# Incorporating point defect generation due to jog formation into the vector density-based continuum dislocation dynamics approach

Peng Lin[1], Vignesh Vivekanandan[1], Benjamin Anglin[2], Clint Geller[2], Anter El-Azab[1]

[1] Purdue University, West Lafayette, IN 47907, USA

[2] Naval Nuclear Laboratory, West Mifflin, PA 15122, USA

Abstract. During plastic deformation of crystalline materials, point defects such as vacancies and interstitials are generated by jogs on moving dislocations. A detailed model for jog formation and transport during plastic deformation was developed within the vector density-based continuum dislocation dynamics framework (Lin and El-Azab, 2020; Xia and El-Azab, 2015). As a part of this model, point defect generation associated with jog transport was formulated in terms of the volume change due to the non-conservative motion of jogs. Balance equations for the vacancies and interstitials including their rate of generation due to jog transport were also formulated. A two-way coupling between point defects and dislocation dynamics was then completed by including the stress contributed by the eigen-strain of point defects. A jog drag stress was further introduced into the mobility law of dislocations to account for the energy dissipation during point defects generation. A number of test problems and a fully coupled simulation of dislocation dynamics and point defect generation and diffusion were performed. The results show that there is an asymmetry of vacancy and interstitial generation due to the different formation energies of the two types of defects. The results also show that a higher hardening rate and a higher dislocation density are obtained when the point defect generation mechanism is coupled to dislocation dynamics.

Keywords: Continuum Dislocation Dynamics; Dislocation jogs; Point defects; Finite Element Method

# 1 Introduction

Defects and their mutual interactions dominate the mechanical behavior of crystalline solids. Point defects such as vacancies are known to interact with dislocations and the study of such interactions is now a classic topic of dislocation theory (Hirth and Lothe, 1982; Hull and Bacon, 2011). Vacancies can affect the performance of crystalline materials in many ways. Vacancy condensation plays an important role in the growth of voids leading to a transition from brittle to ductile fracture (Cuitiño and Ortiz, 1996). Vacancy diffusion can assist dislocation climb, which is an important creep/plastic deformation mechanism in materials such as nickel-based superalloys at high temperature (Gao et al., 2017; Wang et al., 2020; Yang et al., 2015; Yuan et al., 2018). The mobility of dislocations moving via glide can also be affected by vacancies. For example, a so-called vacancy lubrication effect on dislocation motion was discovered (Lu and Kaxiras, 2002) that explains the observed softening in cold-worked high-purity aluminum at low temperatures. Moreover, vacancies seem to play a crucial role in hydrogen embrittlement in metals. Hydrogen-vacancy complexes are stable defect structures that can act as void nucleation sites during plastic deformation (Li et al., 2015). The mobility of hydrogen is greatly impeded by the presence of vacancies (Hayward and Fu, 2013), and dislocations can be locked by hydrogenated vacancies (Xie et al., 2016).

Large numbers of vacancies have been observed in crystalline metals during plastic flow (Klein and Gager, 1966; Seitz, 1952, 1950). It has been pointed out that the average temperature increase due to energy dissipation of dislocation motion is probably not sufficiently high to increase vacancy and interstitial concentrations as a result of thermal effects alone (Seitz, 1952). In fact, theoretical studies have shown that vacancies or interstitials can be generated by purely geometrical means during dislocation motion (Hornstra, 1962; Seeger, 1955; Seitz, 1952, 1950; Zhou et al., 1999; Zsoldos, 1963). The non-conservative motion of dislocations, which is characterized by the motion of dislocations outside of their glide planes, is responsible for the generation of vacancies and interstitials (Hirth and Lothe, 1982; Hull and Bacon, 2011). Different mechanisms can cause the non-conservative motion of dislocations. At high temperature, edge dislocations are able to climb by the addition or deletion of atoms from their cores (Niu et al., 2017). In this case, the dislocation velocity has a component that is perpendicular to the glide plane of the dislocation. Non-conservative motion of dislocations also happen during annihilation of edge dislocations on two closely separated glide planes (Ohashi, 2018; Wang, 2017). This process is similar to dislocation climb, since it can be thought of as if dislocations first climb to the same slip plane and then annihilate. Another important mechanism of point defect generation is the non-conservative motion of jogs (Hornstra, 1962; Hull and Bacon, 2011; Seeger, 1955; Zsoldos, 1963). Dislocations in highly deformed crystals and in crystals under cyclic loading are expected to contain many jogs, which are formed primarily during intersection with other dislocations (Zhou, 1998; Zhou et al., 1999). As a dislocation moves under the action of a stress field, it drags along the jogs that were previously formed on it. The motion of the jogs is non-conservative, since the jog itself does not lie in the glide plane of the dislocation. The non-conservative jog motion may be a major mechanism for vacancy or interstitial generation at low temperature, which is the case we considered in this work. In addition to generating vacancies or interstitials, the presence of jogs can also limit the motion of dislocations (Zhou et al., 1999). Therefore, it is important to study the interactions among dislocations, jogs and point defects, and as such the formation and evolution of jogs should be considered in theoretical models aimed at studying materials containing point defects and dislocations.

Many theoretical models have been used to study the interactions among dislocations, jogs, and point defects in the past. Molecular dynamics (MD) was used to study jog formation (Justo et al., 1997; Zhou, 1998) and vacancy formation (Iyer et al., 2014; Zhou et al., 1999) by dislocation intersection, as well as the energetics of vacancies under different conditions (Gavini, 2008; Iyer et al., 2014; Li et al., 2015). Discrete dislocation dynamics (DDD) is also a practical tool to study dislocation-vacancy interactions (Cui et al., 2018; Po and Ghoniem, 2014; Raabe, 1998). This method was successfully applied to simulate vacancy-assisted dislocation climb in creep behavior of nickel-based superalloy (Gao et al., 2017; Wang et al., 2020), self-climb of dislocation loops by vacancy pipe diffusion (Gao et al., 2011; Pineau et al., 2016; Zhou, 1998). On the other hand, plasticity models at continuum level have also been established to consider the effect of vacancies on plastic deformation of materials (Cuitiño and Ortiz, 1996; Lindgren et al., 2008; Patra and McDowell, 2012; Sahoo, 1984; Yang et al., 2015; Yuan et al., 2018). Even with these successes in understanding the interactions between dislocations and vacancies, there remains two issues to be solved. Due to the computational cost, the length scale and time scale of MD simulations are always limited. The simulation domain can only hold a few dislocation lines. In the meantime, continuum plasticity models usually use a scalar dislocation density quantity to represent dislocation structures, which loses the description of line-like character of dislocations. The line-like character of dislocations is important in modeling the formation and evolution of jogs, and new models are needed to fill the gap between atomistic and continuum descriptions of dislocation-vacancy interactions. Another issue is that the DDD models mentioned above mainly focus on the effect of vacancies on dislocation climb, where thermal activation dominates the behavior of the defect system, while jog formation and vacancy generation by moving jogs are seldom included. As such, when deformation of crystals at low temperature is considered, a proper way for jogs to form with the associated athermal vacancy generation should be included. The above concerns motivate the current work, in which we focus on incorporating the mechanisms of jog formation and vacancy generation into our continuum dislocation dynamics (CDD) framework (Lin and El-Azab, 2020; Xia and El-Azab, 2015).

Continuum dislocation dynamics (CDD) uses density-like field variables to track the evolution of curved dislocation line ensembles. Following the introduction of the dislocation density tensor $\boldsymbol{\alpha}$ by Kröner (1958) and Nye (1953), the time evolution of this tensor was formulated in the form $\dot{\boldsymbol{\alpha}} = \nabla \times (\mathbf{v} \times \boldsymbol{\alpha})$ by Mura (1963) and Kosevich (1965). Such a form is only applicable to families of dislocations of the same Burgers vector and line direction at small resolution since the dislocation velocity field $\mathbf{v}$ is only meaningful at that level. In recent years, several attempts have been made to obtain an average, statistical description of dislocation microstructure evolution. Groma, Zaiser and co-workers (Groma, 1997; Groma et al., 2003; Zaiser et al., 2001) developed statistical approaches in 2D for the evolution of straight edge dislocations; see also the relatively recent works by Kooiman (2014). Arsenlis et al. (2004), Reuber et al. (2014), and Leung et al. (2015) who developed 3D models by including additional line orientation information. However, extending 2D approaches to 3D systems in which dislocations are interconnected curved lines that move perpendicular to their line direction has proven to be challenging. Another approach has been proposed by Hochrainer et al. (2007) to describe the 3D curved dislocation lines by using a higher dimensional phase space containing line direction variables as extra dimensions, so densities can carry additional information about their line direction and curvature. The latter work was motivated by that of El-Azab (2000a, 2000b). A simplified variant of Hochrainer's formulation has been introduced, which considers only low-order moments of the dislocation direction distribution (Sandfeld et al., 2011; Sandfeld and Zaiser, 2015). A further development of this theory has been achieved by defining a hierarchy of evolution equations of the so-called alignment tensors, which contains information on the directional distribution of dislocation density and dislocation curvature (Hochrainer, 2015; Monavari and Zaiser, 2018). The CDD models just mentioned represent important contributions toward describing dislocation transport in crystals while preserving the linear character of dislocations, as in DDD methods. However, the non-conservative effects of dislocation motion are usually neglected, and only dislocations gliding on their slip planes are considered. In order to study dislocation-vacancy interactions, the formation of jogs and its non-conservative motion must be incorporated properly into the CDD framework. Recently, the work done by Hochrainer (2020) showed the vacancy generation by considering the non-conservative motion of dislocations in the CDD framework.

In this paper, the CDD model formulated recently by Xia and El-Azab (2015) is used as a starting point. In this model, the so-called bundle representation of the dislocation density is considered. In this representation, dislocations on each slip system are described by a vector field $\boldsymbol{\rho}^{(k)}$ in such a way that, at the appropriate resolution, the dislocation density has a unique line direction at each point in the crystal. The mesh size required for solving such a model must be sufficiently small to enable the accurate geometric cancellation of dislocations of opposite directions, thus coinciding with the physical annihilation of dislocations. The magnitude of vector $\boldsymbol{\rho}^{(k)}$ gives the scalar dislocation density at each point on the *k*th slip system. The vector field $\boldsymbol{\rho}^{(k)}$ on each slip system evolves via dislocation transport (Xia and El-Azab, 2015), cross slip (Xia et al., 2016), and junction reactions (Lin and El-Azab, 2020). The stress field that drives dislocation transport is fixed by solving an eigen-strain boundary value problem in which the eigen-strain itself is the plastic strain induced by the motion of dislocations. In the present work, the CDD model is augmented by introducing the jog density $\boldsymbol{\rho}_{\mathrm{jog}}^{(k)}$ as an additional dislocation field. A set of equations describing the evolution of the jog density $\boldsymbol{\rho}_{\mathrm{jog}}^{(k)}$ is formulated, in which the rate of jog generation is found from the rate at which dislocations on various slip systems intersect each other. Then the non-conservative motion of the jog density $\boldsymbol{\rho}_{\mathrm{jog}}^{(k)}$ is used to calculate the generation of vacancies and interstitials. The equations describing vacancy and interstitial diffusion and recombination are also established. The effects of both jogs and point defects on the evolution of dislocations are included by adding suitably chosen resistive terms to the dislocation mobility law.

## 2 Vector density-based continuum dislocation dynamics

In the vector density-based CDD approach, the evolution of the dislocation density field is obtained in two steps. The first step is to reduce the classical form of the equation governing the evolution of the dislocation density tensor to that for the corresponding vector density using the dislocation bundle view of the density field. The second step is to build into the resulting evolution equations the rate terms corresponding to cross slip and dislocation reactions. We begin by introducing the definition of the dislocation density tensor, $\boldsymbol{\alpha}$, as given by Kröner (1958) and Nye (1953),

$$\boldsymbol{\alpha} = -\nabla \times \boldsymbol{\beta}^{\mathrm{d}}, \tag{1}$$

with $\boldsymbol{\beta}^{\mathrm{d}}$ being the plastic distortion tensor. Both tensors can be decomposed into slip system contributions,

$$\boldsymbol{\alpha} = \sum_k \boldsymbol{\alpha}^{(k)}, \tag{2}$$

$$\boldsymbol{\beta}^{\mathrm{d}} = \sum_k \boldsymbol{\beta}^{\mathrm{d(k)}}, \tag{3}$$

where *k* is a slip system index. As dislocations move, the plastic distortion will evolve. And its rate can be obtained by Orowan's law,

$$\boldsymbol{\beta}^{\mathrm{d}(k)} = -\mathbf{v}^{(k)} \times \boldsymbol{\alpha}^{(k)} \tag{4}$$

where $\mathbf{v}^{(k)}$ is the dislocation velocity on slip system *k*. Here, we assume the resolution is high enough so that dislocations at a material point have a unique line direction. Hence, the direction of the dislocation velocity $\mathbf{v}^{(k)}$ is taken to be perpendicular to the dislocation line. The dislocation density vector $\boldsymbol{\rho}^{(k)}$ is used to represent the oriented dislocation density at all points, and the relation between the dislocation density vector $\boldsymbol{\rho}^{(k)}$ and the dislocation density tensor $\boldsymbol{\alpha}^{(k)}$ is

$$\boldsymbol{\alpha}^{(k)} = \boldsymbol{\rho}^{(k)} \otimes \mathbf{b}^{(k)}, \tag{5}$$

where $\mathbf{b}^{(k)}$ is the Burgers vector of dislocations on slip system *k*. Combining Eqs. (1) through (5), the evolution equation for dislocation density vector $\boldsymbol{\rho}^{(k)}$ in the absence of reactions and cross slip can be formulated as (Xia and El-Azab, 2015),

$$\dot{\boldsymbol{\rho}}^{(k)} = \nabla \times (\mathbf{v}^{(k)} \times \boldsymbol{\rho}^{(k)}). \tag{6}$$

For multiple slip systems, each slip system has its own dislocation evolution equation in the form of Eq. (6).

In addition to the dislocation glide described by Eq. (6), dislocation reactions among different slip systems also contribute to the dislocation density vector $\boldsymbol{\rho}^{(k)}$. Therefore, additional terms must be added to account for cross slip (Xia et al., 2016; Xia and El-Azab, 2015), collinear annihilation and junction reactions (Lin and El-Azab, 2020) in the evolution equations. For dislocation cross slip from slip system *k* to slip system *l*, the coupling term is defined as (Xia et al., 2016; Xia and El-Azab, 2015)

$$\dot{\boldsymbol{\rho}}_{\text{cs}}^{(k,l)} = i^{(k,l)} c_{\text{cs}}^{(k,l)} (\boldsymbol{\rho}^{(k)} \cdot \tilde{\mathbf{e}}^{(k,l)}) \tilde{\mathbf{e}}^{(k,l)}, \tag{7}$$

In the above, $i^{(k,l)}$ is an indicator, which is unity when the cross slip conditions are satisfied, and zero otherwise (Xia and El-Azab, 2015), $c_{\text{cs}}^{(k,l)}$ is the cross slip rate obtained by coarse graining DDD data (Xia et al., 2016), and $\tilde{\mathbf{e}}^{(k,l)}$ is a unit vector along the intersection of the two cross slip planes. For the two slip systems involved in cross slip, the Burgers vector is also along the intersection. So $\boldsymbol{\rho}^{(k)} \cdot \tilde{\mathbf{e}}$ is the screw component of the dislocation density vector. For collinear annihilation between slip system *k* and slip system *l*, the coupling terms for one time step are defined as (Lin and El-Azab, 2020):

$$\begin{aligned} \Delta\boldsymbol{\rho}_{\text{col}}^{(k)} &= i_{\text{col}}^{(k,l)} \min(|\boldsymbol{\rho}^{(k)} \cdot \tilde{\mathbf{e}}^{(k,l)}|, |\boldsymbol{\rho}^{(l)} \cdot \tilde{\mathbf{e}}^{(k,l)}|) \operatorname{sgn}(\boldsymbol{\rho}^{(k)} \cdot \tilde{\mathbf{e}}^{(k,l)}) \tilde{\mathbf{e}}^{(k,l)} \\ \Delta\boldsymbol{\rho}_{\text{col}}^{(l)} &= i_{\text{col}}^{(k,l)} \min(|\boldsymbol{\rho}^{(k)} \cdot \tilde{\mathbf{e}}^{(k,l)}|, |\boldsymbol{\rho}^{(l)} \cdot \tilde{\mathbf{e}}^{(k,l)}|) \operatorname{sgn}(\boldsymbol{\rho}^{(l)} \cdot \tilde{\mathbf{e}}^{(k,l)}) \tilde{\mathbf{e}}^{(k,l)} \end{aligned}, \tag{8}$$

where $i_{\text{col}}^{(k,l)}$ is an indicator function taking on the value of unity when the annihilation reaction is possible and zero otherwise, and $\tilde{\mathbf{e}}^{(k,l)}$ is the unit vector along the intersection of the slip planes of systems $k$ and $l$ (Lin and El-Azab, 2020). The sign function $\operatorname{sgn}(\cdot)$ is used to ensure that $\Delta\boldsymbol{\rho}_{\text{col}}^{(k)}$ or $\Delta\boldsymbol{\rho}_{\text{col}}^{(l)}$ always form an acute angle with $\boldsymbol{\rho}^{(k)}$ or $\boldsymbol{\rho}^{(l)}$, respectively. Eq. (8) is based on the idea that the rate of collinear annihilation is taken to be the maximum possible rate, fully annihilating the screw component of the smaller density (Lin and El-Azab, 2020). For a glissile junction reaction $\boldsymbol{\rho}^{(k)} + \boldsymbol{\rho}^{(l)} \to \boldsymbol{\rho}^{(m)}$, the reaction rates are defined as follows (Lin and El-Azab, 2020),

$$\dot{\boldsymbol{\rho}}_{\text{g}}^{(kl,m)} = \pm\, i_{\text{g}}^{(k,l)} c_{\text{g}}^{(k,l)} (\boldsymbol{\rho}^{(k)} \cdot \tilde{\mathbf{e}}^{(k,l)})(\boldsymbol{\rho}^{(l)} \cdot \tilde{\mathbf{e}}^{(k,l)}) \tilde{\mathbf{e}}^{(k,l)} \tag{9}$$

where, again, $i_{\text{g}}^{(k,l)}$ is an indicator function that takes on a value of unity when the reaction criterion is satisfied and zero otherwise (Lin and El-Azab, 2020). Here $c_{\text{g}}^{(k,l)}$ is the glissile junction reaction rate, and $\tilde{\mathbf{e}}^{(k,l)}$ is a unit vector along the interaction line of the slip planes of the two reacting slip systems. An energy criterion is used to ensure that $\boldsymbol{\rho}^{(k)} \cdot \tilde{\mathbf{e}}^{(k,l)}$ and $\boldsymbol{\rho}^{(l)} \cdot \tilde{\mathbf{e}}^{(k,l)}$ have the same sign. When they are positive, the positive sign is chosen in Eq. (9). When they are negative, the negative sign is chosen. Coupling these dislocation reactions with dislocation transport, the final form controlling the evolution of dislocations in CDD is

$$\dot{\boldsymbol{\rho}}^{(k)} = \nabla \times (\mathbf{v}^{(k)} \times \boldsymbol{\rho}^{(k)}) - \dot{\boldsymbol{\rho}}_{\mathrm{cs}}^{(k,l)} + \dot{\boldsymbol{\rho}}_{\mathrm{cs}}^{(l,k)} - \dot{\boldsymbol{\rho}}_{\mathrm{col}}^{(k)} - \dot{\boldsymbol{\rho}}_{\mathrm{g}}^{(kl,m)} + \dot{\boldsymbol{\rho}}_{\mathrm{g}}^{(lm,k)} \tag{10}$$

The last two terms in Eq. (10) should include all glissile junctions involving slip system *k*.

# 3 Theoretical development of point defects generation in continuum dislocation dynamics

## 3.1 Point defects generated by non-conservative dislocation motion

There are two types of dislocation motion, conservative and non-conservative (Hirth and Lothe, 1982; Hull and Bacon, 2011). Conservative motion is associated with dislocation glide on the plane containing both its line and Burgers vector. Motion of dislocations outside of this plane has a climb component and is considered non-conservative. The latter type of motion leads to local volume changes in the material, which results in point defect generation (or consumption). In this regard, both vacancies and interstitials can be generated depending on the direction of the non-conservative motion. For example, the formation of vacancies (consumption of interstitials) occurs when edge dislocations climb to extend the extra half-plane (negative climb), while the formation of interstitials (consumption of vacancies) occurs when an edge dislocation climbs to shrink the extra half-plane (positive climb). The number of point defects generated during the non-conservative motion of dislocations can be related to the volume change caused by that motion. If a small dislocation line segment **l** undergoes a small non-conservative displacement **s**, the local volume change is (Hull and Bacon, 2011)

$$\Delta V = \mathbf{b} \cdot (\mathbf{l} \times \mathbf{s}) = \mathbf{s} \cdot (\mathbf{b} \times \mathbf{l}), \tag{11}$$

where **b** is the Burgers vector of the segment. The number of generated point defects compatible with this volume change is given by

$$N = \frac{\Delta V}{\Omega}, \tag{12}$$

where $\Omega$ is the volume of an atom. The sign of $\Delta V$ determines the type of point defect. When the Burgers vector is defined by right-hand/finish-start convention, $\mathbf{l} \times \mathbf{b}$ always points to the extra half plane of the edge dislocation. Combined with the definition of negative and positive climb, vacancy generation occurs when $\Delta V$ is positive and interstitial generation occurs when $\Delta V$ is negative.

Eqs. (11) and (12) link the point defect generation to the motion of a discrete dislocation line. In CDD, dislocations are represented by a dislocation density vector $\boldsymbol{\rho}$. Point defects can be measured by their concentration $c_d$. As such, the rate of generation of point defects due to non-conservative dislocation motion can be expressed in terms of the dislocation density vector as

$$\dot{c}_d = \frac{\mathbf{b}\cdot(\boldsymbol{\rho}\times\mathbf{v})}{\Omega} \tag{13}$$

where **v** is the dislocation velocity. Eq. (13) is valid for all dislocation characters, pure edge, pure screw, or mixed. For a screw dislocation, **b** is parallel to $\boldsymbol{\rho}$, so Eq. (13) yields zero rate of defect production, meaning the motion of pure screw dislocation is always conservative and no point defects will be generated.

When the temperature is too low, thermally activated dislocation climb is unlikely to occur (Hirth and Lothe, 1982; Hull and Bacon, 2011). However, two mechanisms of non-conservative dislocation motion that may operate at any temperature (not requiring thermal activation) are the edge dipole annihilation (Aslanides and Pontikis, 2000; Brinckmann et al., 2011) and jog motion (Hornstra, 1962; Seeger, 1955; Seitz, 1952; Zsoldos, 1963), see Figure 1. A pair of opposite edge dislocations on closely separated glide planes can approach each other, annihilate and generate point defects. This process can be considered as a climb of one dislocation to the slip plane of the other dislocation resulting in mutual annihilation. Another athermal deformation mechanism is the movement of a jogged screw dislocation (Hornstra, 1962; Seeger, 1955; Seitz, 1952; Zsoldos, 1963). Jogs are steps on the dislocation line that move it from one atomic slip plane to another. The jog on a screw dislocation has edge character. When a screw dislocation glides under an applied stress, it drags the jogs along. Such a motion requires jog climb, which generates a point defect trail behind the jog. Therefore, jogs can be considered as sources for point defects. Eq. (13) is valid for both the dipole annihilation and jog drag mechanisms. The current work mainly focuses on point defect generation by moving jogs. In what follows, a jog density $\boldsymbol{\rho}_{\text{jog}}$ will be introduced and used to replace $\boldsymbol{\rho}$ in Eq. (13) in order to determine the defect generation rate due to jog motion.

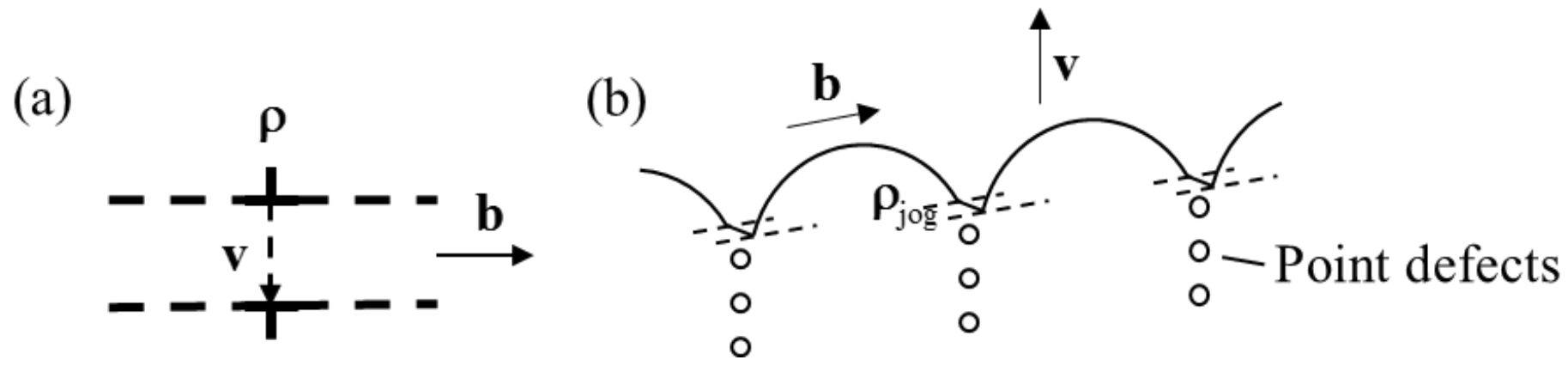

Figure 1. Two mechanisms of non-conservative dislocation motion at room temperature. (a) Annihilation of an edge dipole on closely separated glide planes. (b) Motion of a jogged screw dislocation.

## 3.2 Jogs formed by intersection of dislocations

Jogs on dislocations in deforming crystals are formed by intersections of dislocations, as shown in Figure 2. When one dislocation is cut by another dislocation a jog forms at the intersection. The jog segment is equal to the Burgers vector of the cutting dislocation. In Figure 2, a dislocation segment $\mathbf{l}^{(1)}$ moves with velocity $\mathbf{v}^{(1)}$ and a second dislocation segment $\mathbf{l}^{(2)}$ is stationary. The corresponding Burgers vectors are $\mathbf{b}^{(1)}$ and $\mathbf{b}^{(2)}$ and the arrows indicate their directions.

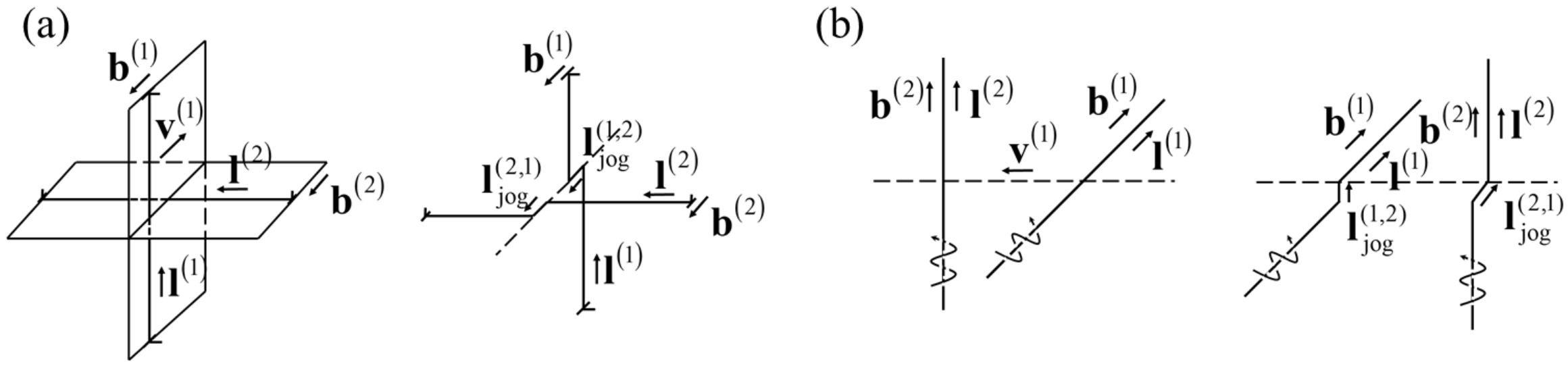

Figure 2. Jogs formed by intersection of dislocations. (a) Intersection of edge dislocations with parallel Burgers vector. (b) Intersection of two right-handed screw dislocations.

From Figure 2, it is easy to see that the jog formed on each dislocation line upon dislocation intersection can be expressed as

$$\mathbf{l}_{\text{jog}}^{(1,2)} = \mathbf{b}^{(2)}, \quad \mathbf{l}_{\text{jog}}^{(2,1)} = \mathbf{b}^{(1)}, \tag{14}$$

where $\mathbf{l}_{\text{jog}}^{(k,l)}$ denotes the jog segment on dislocation *k* cut by dislocation *l*. Eq. (14) is valid for the case shown in Figure 2. However, there can be a negative sign in Eq. (14). Imagining a different case from Figure 2 (a) with $\mathbf{l}^{(1)}$ moving in the opposite direction, $\mathbf{l}_{\text{jog}}^{(1,2)}$ will be equal to $-\mathbf{b}^{(2)}$. Whether the jog formed is $+\mathbf{b}^{(2)}$ or $-\mathbf{b}^{(2)}$ is determined by the direction of three vectors, the line direction of the two dislocations, $\mathbf{l}^{(1)}$ and $\mathbf{l}^{(2)}$, and the relative velocity of the two dislocations, $(\mathbf{v}^{(2)}-\mathbf{v}^{(1)})$, i.e., the direction of the relative displacement. A triple vector product of these three vectors can be used to calculate the sign (Hornstra, 1962; Zsoldos, 1963). So, instead of Eq. (14), the jog segments in the general case should be

$$\mathbf{l}_{\text{jog}}^{(k,l)} = \text{sgn}((\mathbf{v}^{(l)}-\mathbf{v}^{(k)})\cdot(\mathbf{l}^{(k)}\times\mathbf{l}^{(l)}))\,\mathbf{b}^{(l)}, \tag{15}$$

where sgn($x$) is the sign function, and $k$ and $l$ are slip system indices.

Eq. (15) describes the jogs formed by intersection of two dislocation lines. To incorporate this result into CDD, a continuum description of the jogs in terms of dislocation density vectors must be established. As discussed earlier, the dislocation density vector $\boldsymbol{\rho}^{(k)}$ locally represents a bundle of dislocations with the same line direction. Thus when two dislocation bundles cut each other at a material point, all jog segments formed on one dislocation bundle have the same length and direction, which can be obtained by Eq. (15). Then a jog density $\boldsymbol{\rho}_{\text{jog}}^{(k,l)}$ can be defined as the oriented length of the jog segments per unit volume. This jog density is given by

$$\dot{\boldsymbol{\rho}}_{\text{jog}}^{(k,l)} = \frac{N_{\text{jog}}^{(k,l)}}{V\Delta t}\mathbf{l}_{\text{jog}}^{(k,l)} \tag{16}$$

with $N_{\text{jog}}^{(k,l)}$ being the number of jogs formed on slip system $k$ by slip system $l$ in a control volume $V$ during time $\Delta t$. We now explain this formula and fix $N_{\text{jog}}^{(k,l)}$ in terms of $\boldsymbol{\rho}^{(k)}$ and $\boldsymbol{\rho}^{(l)}$. Consider a control volume element $V$ as shown in Figure 3(a), with the $x$-axis taken along the dislocation density vector $\boldsymbol{\rho}^{(1)}$ and corresponding edge of $l_x$, $y$-axis along $\boldsymbol{\rho}^{(2)}$ and corresponding edge of $l_y$, and the $z$-axis along the relative velocity $(\mathbf{v}^{(2)}-\mathbf{v}^{(1)})$ with the corresponding edge $l_z$. The volume $V$ of this control volume is the magnitude of the triple product of the three vectors forming its edges,

$$V = \left| l_z \frac{(\mathbf{v}^{(2)}-\mathbf{v}^{(1)})}{\|\mathbf{v}^{(2)}-\mathbf{v}^{(1)}\|}\cdot l_x \frac{\boldsymbol{\rho}^{(1)}}{\|\boldsymbol{\rho}^{(1)}\|}\times l_y \frac{\boldsymbol{\rho}^{(2)}}{\|\boldsymbol{\rho}^{(2)}\|}\right| = l_x l_y l_z \frac{|(\mathbf{v}^{(2)}-\mathbf{v}^{(1)})\cdot(\boldsymbol{\rho}^{(1)}\times\boldsymbol{\rho}^{(2)})|}{\|\mathbf{v}^{(2)}-\mathbf{v}^{(1)}\|\cdot\|\boldsymbol{\rho}^{(1)}\|\cdot\|\boldsymbol{\rho}^{(2)}\|}. \tag{17}$$

In Figure 3(b) and Figure 3(c), the red and green arrows represent the two intersecting dislocation bundles. By the definition of dislocation density, the number of dislocations in the bundles can be calculated as

$$N^{(1)} = \frac{\|\boldsymbol{\rho}^{(1)}\|V}{l_x}, \quad N^{(2)} = \frac{\|\boldsymbol{\rho}^{(2)}\|V}{l_y} \tag{18}$$

For a given time increment, $\Delta t$, the displacement of $\boldsymbol{\rho}^{(2)}$ relative to $\boldsymbol{\rho}^{(l)}$ is $(\mathbf{v}^{(2)} - \mathbf{v}^{(1)})\Delta t$. Here we assume that the dislocations are uniformly distributed in this small control volume. Each dislocation in the $\boldsymbol{\rho}^{(2)}$ field will then intersect with $N^{(1)} \dfrac{\| \mathbf{v}^{(2)} - \mathbf{v}^{(1)} \| \Delta t}{l_z}$ dislocations in the $\boldsymbol{\rho}^{(1)}$ field, so the total number of jogs formed on $\boldsymbol{\rho}^{(1)}$ will be

$$N_{\text{jog}}^{(1,2)} = N^{(1)} N^{(2)} \frac{\| \mathbf{v}^{(2)} - \mathbf{v}^{(1)} \| \Delta t}{l_z} = \frac{V^2 \Delta t}{l_x l_y l_z} \| \mathbf{v}^{(2)} - \mathbf{v}^{(1)} \| \cdot \| \boldsymbol{\rho}^{(1)} \| \cdot \| \boldsymbol{\rho}^{(2)} \| . \tag{19}$$

By substituting Eqs. (15), (17) and (19) into Eq. (16), and replacing 1 and 2 by $k$ and $l$, we reach

$$\dot{\boldsymbol{\rho}}_{\text{jog}}^{(k,l)} = (\mathbf{v}^{(l)} - \mathbf{v}^{(k)}) \cdot (\boldsymbol{\rho}^{(k)} \times \boldsymbol{\rho}^{(l)}) \mathbf{b}^{(l)} , \tag{20}$$

which is the continuum rate form for jog formation of the density $\boldsymbol{\rho}^{(k)}$ by intersection with the density $\boldsymbol{\rho}^{(l)}$ that is compatible with the CDD formulation.

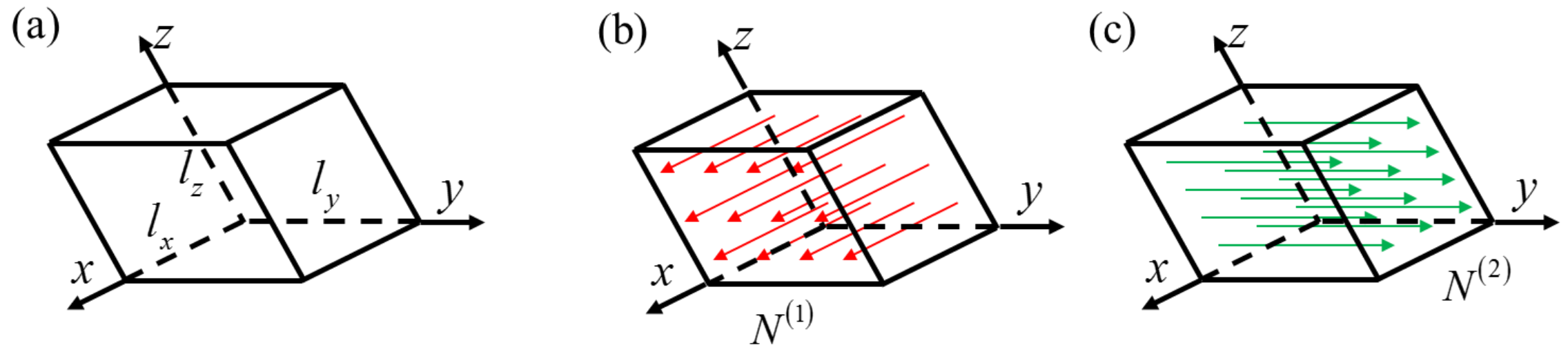

Figure 3. (a) Representative volume used to calculate the number of jogs at a material point. The axes are along the unit vectors $\mathbf{e}_x = \boldsymbol{\rho}^{(1)} / \| \boldsymbol{\rho}^{(1)} \|$, $\mathbf{e}_y = \boldsymbol{\rho}^{(2)} / \| \boldsymbol{\rho}^{(2)} \|$ and $\mathbf{e}_z = (\mathbf{v}^{(2)} - \mathbf{v}^{(1)}) / \| \mathbf{v}^{(2)} - \mathbf{v}^{(1)} \|$. (b) Number of dislocation lines in dislocation bundle $\boldsymbol{\rho}^{(1)}$ is $N^{(1)}$; (c) Number of dislocation lines in dislocation bundle $\boldsymbol{\rho}^{(2)}$ is $N^{(2)}$.

### 3.3 Transport equations for jog densities

As discussed in Section 3.1, the non-conservative motion of jogs leads to point defect generation at a rate proportional to the instantaneous density of jogs. As such, establishing the evolution equations of jog density is important to connecting plastic deformation and point defect generation. Generally speaking, as jogs are small steps (only a Burgers vector height) on the dislocation line, the jog density $\boldsymbol{\rho}_{\text{jog}}^{(k,l)}$ is significantly smaller than the glide dislocation density $\boldsymbol{\rho}^{(k)}$. Therefore, although the existence of jogs may change the average direction of the dislocation lines, the jogs are treated as "particles" tied to the parent dislocations and the dislocations are assumed to remain smooth and planar as far as glide motion is concerned. The concentration of a certain type of jog $c_{\text{jog}}^{(k,l)}$ can be defined as the number of jogs per volume, which is defined by the relation $\boldsymbol{\rho}_{\text{jog}}^{(k,l)} = c_{\text{jog}}^{(k,l)}\mathbf{b}^{(l)}$. If jog migration along the dislocation line is ignored, jogs will move with the same velocity as dislocations. That is, all type of jogs $c_{\text{jog}}^{(k,l)}$ on dislocation $\boldsymbol{\rho}^{(k)}$ have velocity $\mathbf{v}^{(k)}$ of the line. Hence the evolution of the jog concentration $c_{\text{jog}}^{(k,l)}$ follows the transport equation

$$\dot{c}_{\text{jog}}^{(k,l)} + \nabla\cdot(c_{\text{jog}}^{(k,l)}\mathbf{v}^{(k)}) = (\mathbf{v}^{(l)} - \mathbf{v}^{(k)})\cdot(\boldsymbol{\rho}^{(k)}\times\boldsymbol{\rho}^{(l)}) \tag{21}$$

The right hand side of Eq. (21) is a source term for jogs, see Eq. (20). The point defect generation rate is the sum over all jogs from Eq. (13),

$$\dot{c}_d = \sum_k\sum_l \frac{\mathbf{b}^{(k)}\cdot(\boldsymbol{\rho}_{\text{jog}}^{(k,l)}\times\mathbf{v}^{(k)})}{\Omega} = \sum_k \frac{\mathbf{b}^{(k)}\cdot(\sum_l \boldsymbol{\rho}_{\text{jog}}^{(k,l)}\times\mathbf{v}^{(k)})}{\Omega} \tag{22}$$

It can be seen from Eq. (22) that the evolution of each type of jog is not necessarily needed to calculate the point defect generation rate. What matters is the sum of the jog densities on a specific dislocation since they all have the same velocity and Burgers vector. This approximation takes the vector sum of the jogs on a dislocation line, which is valid at small resolution, and is indeed the case for our CDD model. We now define $\boldsymbol{\rho}_{\text{jog}}^{(k)} = \sum_l \boldsymbol{\rho}_{\text{jog}}^{(k,l)}$ to be the vector jog density on dislocations belonging to slip system $k$. The evolution equations of $\boldsymbol{\rho}_{\text{jog}}^{(k)}$ can be derived using the definition $\boldsymbol{\rho}_{\text{jog}}^{(k,l)} = c_{\text{jog}}^{(k,l)}\mathbf{b}^{(l)}$ and Eq. (21),

$$\dot{\boldsymbol{\rho}}_{\text{jog}}^{(k)} + \nabla\cdot(\mathbf{v}^{(k)}\otimes\boldsymbol{\rho}_{\text{jog}}^{(k)}) = \sum_l (\mathbf{v}^{(l)} - \mathbf{v}^{(k)})\cdot(\boldsymbol{\rho}^{(k)}\times\boldsymbol{\rho}^{(l)})\mathbf{b}^{(l)} \tag{23}$$

Then the point defect generation rate can be written in terms of $\boldsymbol{\rho}_{\text{jog}}^{(k)}$ as

$$\dot{c}_d = \frac{1}{\Omega}\sum_k \mathbf{b}^{(k)}\cdot(\boldsymbol{\rho}_{\text{jog}}^{(k)}\times\mathbf{v}^{(k)}) \tag{24}$$

Eqs. (23) and (24), respectively, describe the jog density evolution and the rate of point defect generation by jogs on dislocation lines during plastic deformation.

## 3.4 Point defect balance equations

Vacancies and interstitials can be lost either through recombination or by reaction with a defect sink such as a dislocation, a grain boundary, or a precipitate. The local change in defect concentration of the defect species is the net result of the local production rate, reactions with other species, and diffusion. In our model, the source for point defect production is the non-conservative motion of dislocation jogs as described by Eq. (24). In the current model, and for the sake of an initial implementation, only recombination of vacancies and interstitials is considered besides generation by jogs and diffusion. The recombination rate is given by (Was, 2016)

$$\dot{c}_{\mathrm{v}} = \dot{c}_{\mathrm{i}} = -K_{\mathrm{iv}} c_{\mathrm{i}} c_{\mathrm{v}} \tag{25}$$

where $K_{\mathrm{iv}}$ is vacancy-interstitial recombination rate constant, which is given in terms of the diffusivities of vacancies and interstitials as follows:

$$K_{\mathrm{iv}} = 4\pi r_{\mathrm{iv}} (D_{\mathrm{i}} + D_{\mathrm{v}}) \approx 4\pi r_{\mathrm{iv}} D_{\mathrm{i}}\,, \tag{26}$$

with $D_{\mathrm{i}}$ and $D_{\mathrm{v}}$ being the diffusivities of interstitials and vacancies, respectively. $r_{\mathrm{iv}}$ is the interaction radius. The random walk of defects in the lattice give rise to diffusive fluxes for vacancies, $\mathbf{J}_{\mathrm{v}}$, and interstitial, $\mathbf{J}_{\mathrm{i}}$, which are expressed in terms of the concentration and pressure gradients in the form

$$\mathbf{J}_{\mathrm{v}} = -D_{\mathrm{v}} (\nabla c_{\mathrm{v}} - \frac{c_{\mathrm{v}} \Delta\Omega_{\mathrm{v}} \nabla p}{k_{\mathrm{B}} T}) \quad \text{and} \quad \mathbf{J}_{\mathrm{i}} = -D_{\mathrm{i}} (\nabla c_{\mathrm{i}} - \frac{c_{\mathrm{i}} \Delta\Omega_{\mathrm{i}} \nabla p}{k_{\mathrm{B}} T}) \tag{27}$$

where the pressure $p = -\sigma_{ii}/3$, with $\sigma_{ii}$ being the trace of the stress tensor. Combining the point defect source from jog motion, Eq. (24), and recombination of vacancies and interstitials Eq. (25), the balance equations for vacancies and interstitials are obtained,

$$\begin{aligned} \dot{c}_{\mathrm{v}} &= \nabla \cdot (D_{\mathrm{v}} \nabla c_{\mathrm{v}} - \frac{D_{\mathrm{v}} c_{\mathrm{v}} \Delta\Omega_{\mathrm{v}} \nabla p}{k_{\mathrm{B}} T}) + \dot{c}_{\mathrm{v,s}} - k_{\mathrm{iv}} c_{\mathrm{i}} c_{\mathrm{v}}\,, \\ \dot{c}_{\mathrm{i}} &= \nabla \cdot (D_{\mathrm{i}} \nabla c_{\mathrm{i}} - \frac{D_{\mathrm{i}} c_{\mathrm{i}} \Delta\Omega_{\mathrm{i}} \nabla p}{k_{\mathrm{B}} T}) + \dot{c}_{\mathrm{i,s}} - k_{\mathrm{iv}} c_{\mathrm{i}} c_{\mathrm{v}}\,. \end{aligned} \tag{28}$$

In the above equations, $\dot{c}_{\mathrm{v,s}}$ and $\dot{c}_{\mathrm{i,s}}$ are the generation rates of vacancies and interstitials by jogs. It is to be noted that, at any given point in space, jogs can either generate vacancies or interstitials but not both.

# 4 Coupling dislocation dynamics and mechanics

## 4.1 Stress field stemming from lattice defects system

For a crystal with lattice defects, the stress field includes contributions due to the boundary conditions and the internal defects fields. In order to determine the stress state, the kinematics of crystal deformation in terms of defect contributions first is fixed. The crystal distortion $\boldsymbol{\beta}$ is related to the displacement field $\mathbf{u}$ by

$$\boldsymbol{\beta} = \nabla \mathbf{u} . \tag{29}$$

This distortion is decomposed into four parts (Po and Ghoniem, 2014),

$$\boldsymbol{\beta} = \boldsymbol{\beta}^{\mathrm{e}} + \boldsymbol{\beta}^{\mathrm{d}} + \boldsymbol{\beta}^{\mathrm{v}} + \boldsymbol{\beta}^{\mathrm{i}} \tag{30}$$

where $\boldsymbol{\beta}^{\mathrm{e}}$ is elastic distortion and $\boldsymbol{\beta}^{\mathrm{d}}$, $\boldsymbol{\beta}^{\mathrm{v}}$, and $\boldsymbol{\beta}^{\mathrm{i}}$ are the inelastic distortions induced by dislocations, vacancies, and interstitials, respectively. The dislocation distortion $\boldsymbol{\beta}^{\mathrm{d}}$ is updated by the method of field dislocation mechanics (Acharya and Roy, 2006; Roy and Acharya, 2006) where it is expressed in the form $\boldsymbol{\beta}^{\mathrm{d}} = \nabla \mathbf{z} - \boldsymbol{\chi}$, with $\nabla \mathbf{z}$ and $\boldsymbol{\chi}$ being the compatible and incompatible parts of $\boldsymbol{\beta}^{\mathrm{d}}$, respectively. These two components of the plastic distortion are governed by the following boundary value problems:

$$\begin{cases} \nabla \times \boldsymbol{\chi} = \sum_k \boldsymbol{\rho}^{(k)} \otimes \mathbf{b}^{(k)} & \text{in } V \\ \nabla \cdot \boldsymbol{\chi} = \mathbf{0} & \text{in } V \\ \mathbf{n} \cdot \boldsymbol{\chi} = \mathbf{0} & \text{on } \partial V \end{cases} \tag{31}$$

and,

$$\begin{cases} \nabla \cdot \nabla \dot{\mathbf{z}} = \nabla \cdot \sum_k (-\mathbf{v}^{(k)} \times \boldsymbol{\rho}^{(k)} \otimes \mathbf{b}^{(k)}) & \text{in } V \\ \mathbf{n} \cdot \nabla \dot{\mathbf{z}} = \mathbf{n} \cdot \sum_k (-\mathbf{v}^{(k)} \times \boldsymbol{\rho}^{(k)} \otimes \mathbf{b}^{(k)}) & \text{on } \partial V \\ \dot{\mathbf{z}} = \dot{\mathbf{z}}_o \,(\text{arbitrary value}) & \text{at one point in } V \end{cases} . \tag{32}$$

Here, $V$ is the simulation domain with boundary $\partial V$. It has been shown that updating $\boldsymbol{\beta}^{\mathrm{d}}$ by field dislocation mechanics is more accurate than directly integrating Orowan's equation (Lin et al., 2020). To calculate the eigen-distortions due to vacancies and interstitials, point defects are considered as spherical inclusions in the crystal (Cai et al., 2014), that is, inserting defects into the lattice results in volumetric expansion or contraction. Suppose the volume of the crystal increases by $\Delta\Omega_{\mathrm{i}}$ due to one interstitial, the eigen-distortion field of interstitials can be related to its volume concentration by (Hull and Bacon, 2011; Po and Ghoniem, 2014)

$$\boldsymbol{\beta}^{\mathrm{i}} = \frac{1}{3} c_{\mathrm{i}} \Delta\Omega_{\mathrm{i}} \mathbf{I}, \tag{33}$$

where $c_{\mathrm{i}}$ is the volumetric concentration of interstitials. Similarly, if the volume changes by $\Delta\Omega_{\mathrm{v}}$ due to one vacancy, the corresponding eigen-distortion is given by

$$\boldsymbol{\beta}^{\mathrm{v}} = \frac{1}{3} c_{\mathrm{v}} \Delta\Omega_{\mathrm{v}} \mathbf{I}, \tag{34}$$

where $c_{\mathrm{v}}$ is the volume concentration of vacancies. It should be pointed out that $\Delta\Omega_{\mathrm{v}}$ is negative for vacancy.

With all eigen-distortions known in terms of the corresponding densities of defects, the stress field $\boldsymbol{\sigma}$ can be calculated by a standard Cauchy equilibrium equation,

$$\begin{cases} \nabla \cdot \boldsymbol{\sigma} = \mathbf{0} & \text{in } V \\ \boldsymbol{\sigma} = \mathbf{C} : (\nabla \mathbf{u} - \boldsymbol{\beta}^{\mathrm{d}} - \boldsymbol{\beta}^{\mathrm{v}} - \boldsymbol{\beta}^{\mathrm{i}})_{\mathrm{sym}} & \text{in } V \\ \mathbf{u} = \bar{\mathbf{u}} & \text{on } \partial V_u \\ \mathbf{n} \cdot \boldsymbol{\sigma} = \bar{\mathbf{t}} & \text{on } \partial V_\sigma \end{cases} \tag{35}$$

where $\partial V_u$ and $\partial V_\sigma$ are the parts of the boundaries corresponding to displacement and traction constraints, respectively, and $\mathbf{C}$ is the elastic stiffness tensor. By solving Eq. (35), the stress field combining both boundary conditions and defect effects can be found and used to calculate the velocity of dislocations and the pressure gradient terms in the diffusion equations.

## 4.2 Dislocation mobility law and jog drag

The dislocation velocity is required to close the dislocation transport-reaction equations (10), the jog transport equations (23), and the point defect generation rate expression (24). The dislocation velocity $\mathbf{v}^{(k)}$ is expressed in the form

$$\mathbf{v}^{(k)} = v^{(k)} \boldsymbol{\eta}^{(k)} \tag{36}$$

with $v^{(k)}$ being the scalar velocity and $\boldsymbol{\eta}^{(k)}$ a unit vector in the direction of dislocation motion, which is determined in terms of the slip plane normal $\mathbf{m}^{(k)}$ and the dislocation line direction $\boldsymbol{\xi}^{(k)} = \boldsymbol{\rho}^{(k)} / \rho^{(k)}$ by the following expression

$$\boldsymbol{\eta}^{(k)} = \mathbf{m}^{(k)} \times \boldsymbol{\xi}^{(k)}. \tag{37}$$

The scalar velocity $v^{(k)}$ is assumed to depend linearly on the resolved shear, $\tau^{(k)}$,

$$v^{(k)} = \mathrm{sgn}(\tau^{(k)})\frac{b}{B}[|\tau^{(k)}| - (\tau_0^{(k)} + \tau_{\mathrm{p}}^{(k)} + \tau_{\mathrm{jd}}^{(k)})] \tag{38}$$

where $\mathrm{sgn}(\cdot)$ is the sign function, $b$ is the magnitude of Burgers vector, $B$ is the drag coefficient, and $\tau_0^{(k)}$, $\tau_{\mathrm{p}}^{(k)}$ and $\tau_{\mathrm{jd}}^{(k)}$, respectively, are contributions to the friction stress discussed below. In the above expression, $b\tau^{(k)}$ corresponds to the magnitude of the Peach-Koehler force (Peach and Koehler, 1950), and the resolved shear stress itself is given by

$$\tau^{(k)} = \mathbf{s}^{(k)} \cdot \boldsymbol{\sigma} \cdot \mathbf{m}^{(k)} \tag{39}$$

where $\mathbf{s}^{(k)} = \mathbf{b}^{(k)} / b$ is the unit slip direction. The resolved shear stress accounts for the long-range interactions between dislocations and the dislocation-defect interaction, as well as the boundary effects.

The friction stress $\tau_0^{(k)}$ is the threshold stress for dislocation motion (Hirth and Lothe, 1982; Hull and Bacon, 2011), while $\tau_{\mathrm{p}}^{(k)}$ is the resistance caused by short-range interactions with dislocation junctions (Deng and El-Azab, 2010; El-Azab, 2000a; Hochrainer, 2016; Sandfeld and Zaiser, 2015). This stress is assumed here to take the form of a Taylor hardening law of the form (Devincre et al., 2006; Franciosi et al., 1980; Kubin et al., 2008):

$$\tau_{\mathrm{p}}^{(k)} = \mu b \sqrt{a^{kl} \rho^{(l)}} \tag{40}$$

with $\mu$ being the shear modulus and $a^{kl}$ an interaction matrix representing the average strength of the mutual interactions between slip systems $k$ and $l$. The number of distinct interaction coefficients between 12 mutually interacting slip systems in a FCC crystal is reduced to only six due to symmetry. These are the self, coplanar, and collinear interactions and glissile, Lomer, and Hirth junction (Devincre et al., 2006; Kubin et al., 2008; Madec, 2003). Finally, $\tau_{\mathrm{jd}}^{(k)}$ is the drag stress due to jogs on dislocations. The work done against this drag stress corresponds to the energy used to produce point defects during jog motion. The point defect generation rate is shown in Eq. (24), so the rate of energy used in the process by jogs on dislocations on a given slip system can be easily obtained as

$$\dot{E}^{(k)} = \frac{E_{\mathrm{d}}}{\Omega} |\mathbf{b}^{(k)} \cdot (\boldsymbol{\rho}_{\mathrm{jog}}^{(k)} \times \mathbf{v}^{(k)})| \tag{41}$$

where $E_{\mathrm{d}}$ is the point defect formation energy, and the defect itself can be either vacancy or interstitial based on the sign of $\mathbf{b}^{(k)} \cdot (\boldsymbol{\rho}_{\mathrm{jog}}^{(k)} \times \mathbf{v}^{(k)})$. The rate of work done against the jog drag stress $\tau_{\mathrm{jd}}^{(k)}$ is (Hirth and Lothe, 1982; Hull and Bacon, 2011)

$$\dot{W}^{(k)} = \tau_{\mathrm{jd}}^{(k)} b^{(k)} \rho^{(k)} v^{(k)} . \tag{42}$$

By equating $\dot{E}^{(k)}$ with $\dot{W}^{(k)}$ from the last two expressions, the jog drag stress is found to have the form

$$\tau_{\mathrm{jd}}^{(k)} = \frac{E_{\mathrm{d}}}{\Omega} \cdot \left| \frac{\mathbf{b}^{(k)}}{b} \cdot \frac{\boldsymbol{\rho}_{\mathrm{jog}}^{(k)}}{\rho^{(k)}} \times \frac{\mathbf{v}^{(k)}}{v^{(k)}} \right| . \tag{43}$$

Substituting Eqs. (36) and (37) into Eq. (43), leads to the following final form of the jog drag stress

$$\tau_{\mathrm{jd}}^{(k)} = \frac{E_{\mathrm{d}}}{(\rho^{(k)})^2 \Omega} \cdot \left| \mathbf{s}^{(k)} \cdot (\boldsymbol{\rho}_{\mathrm{jog}}^{(k)} \times (\mathbf{m}^{(k)} \times \boldsymbol{\rho}^{(k)})) \right| . \tag{44}$$

As mentioned earlier, a jog can either generate vacancies or interstitials based upon its character and direction of motion. However, the formation energy of an interstitial is larger than the formation energy of a vacancy. As such, for the same dislocation speed, a jog generating interstitials will move slower and generate less interstitials than one generating vacancies.

# 5 Numerical implementation

The finite element method (FEM) is used to solve coupled crystal mechanics, dislocation transport, jog transport, and point defect diffusion problems. The standard Galerkin finite element method (SGFEM) (Belytschko et al., 2013) and the least squares finite element method (LSFEM) (Jiang, 2013) were both used in solving the coupled problem. The compatible part of plastic distortion of dislocations, stress equilibrium, and point defect diffusion problems was solved by the SGFEM method. The incompatible part of the plastic distortion of dislocations, dislocation transport, and the jog transport equations problems was solved by the LSFEM method, which yields stable and accurate solution for the div-curl type and convective transport equations (Jiang, 2013; Varadhan et al., 2006). Details of the numerical formulations can be found in our earlier work (Lin et al., 2020; Lin and El-Azab, 2020; Xia and El-Azab, 2015).

At a given time step, the dislocation density and point defect concentrations from the previous time step are used to update the eigen-distortions of these fields. The stress field is then computed, from which the dislocation velocity follows. The mobility law is employed in which the resolved shear stress from the stress field and the jog drag stress are used. Then the evolution of dislocation densities is computed. The intersection of the dislocation densities on various slip systems are considered as source terms for the jog evolution equations. The non-conservative motion of dislocation jogs is calculated to generate point defects, which is then used to solve the diffusion equations of point defects. As all the variables are updated, the simulation proceeds to a new time step.

The mesh for the FEM is a hybrid mesh with pyramid and tetrahedron elements (Xia and El-Azab, 2015). This mesh enables us to obtain accurate results of dislocation transport in FCC crystals, since the slip planes can be represented by the faces of the finite elements exactly. The time step $\Delta t$ is adapted using the maximum dislocation velocity, $v_{\max}$, over the simulation domain as follows

$$\frac{v_{\max}\Delta t}{l_{\text{mesh}}} = C. \tag{45}$$

with $l_{\text{mesh}}$ being the mesh size and $C$ the Courant number, which is taken here to be 0.45.

# 6 Results and discussion

To verify the coupled model established in the previous sections, simple test simulations were performed, which illustrate how point defects are generated by jogs on moving dislocations. Bulk simulations of a FCC crystal under uniaxial loading were then performed to elucidate the differences in the mechanical behavior of the crystal when jog drag and the point defect generation mechanism is taken into consideration.

## 6.1 Jogs and vacancies generated by two intersecting dislocation loop bundles

Initially, there are two dislocation bundles in the form of loops, which we will call loops for brevity, placed on two different slip planes in a $2\mu\text{m}\times 2\mu\text{m}\times 6.364\mu\text{m}$ simulation volume, as shown in Figure 4. Over a cross section of the loop the density exhibits a gaussian distribution in the radial direction in the slip plane and in the direction normal to the slip plane. The red loop is on a slip plane with normal along the *x*-axis and with Burgers vector along the *y*-axis, while the green loop is on a slip plane with normal along the *y*-axis and with Burgers vector along the *x*-axis. In this test, the slip systems do not coincide with those of a FCC crystal. They are rather chosen to make the analysis easier in this test problem. The dislocation velocity for both loops is assigned a constant value of $0.03\mu m/ns$ such that they expand to intersect each other. Periodic boundary conditions are also employed.

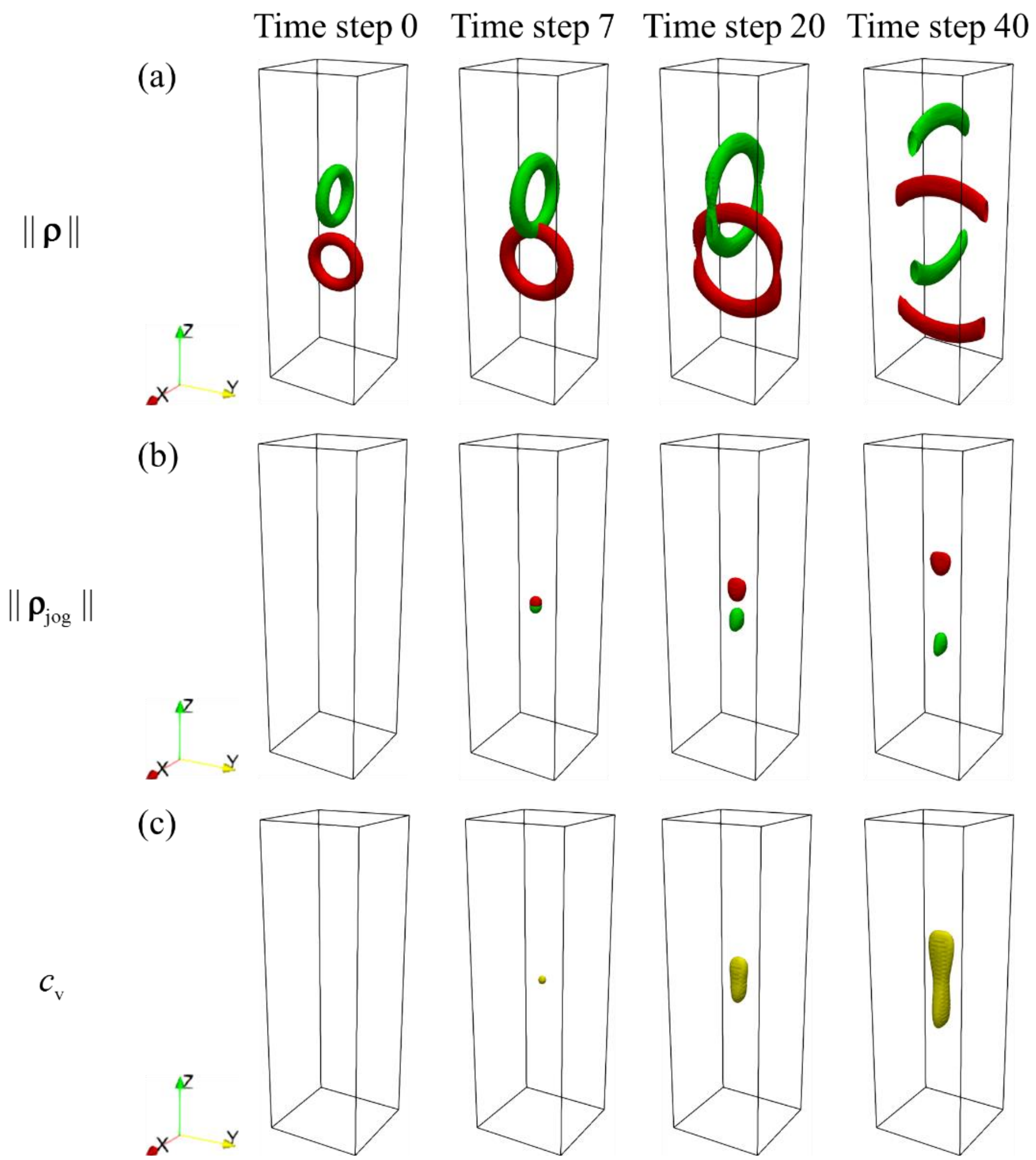

Figure 4. Evolution of (a) dislocation densities, (b) jog densities and (c) vacancy concentration as two dislocation loop bundles expanding and intersecting with each other. Contour surfaces are used in these figures to show the results in 3D.

Figure 4 shows the evolution of the dislocation density, jog density, and point defect concentration. Before the two loops intersected with each other there were no jogs and no vacancies in the domain. At the onset of intersection, jogs start to form at the places where the two loops intersect. As more of the bundles intersect, the region with jogs becomes larger. After the two dislocation loops pass each other, no more jogs form and the existing jogs move together with the corresponding dislocations. In Figure 4 (a) and (b), the same color is used for dislocations and jogs on the same slip system. The intersecting parts of the two dislocation bundles have screw character, and the jogs on them have edge character. The motion of screw dislocations with jogs having edge character will result in point defect generation during slip (Hull and Bacon, 2011). As can be seen in Figure 4 (c), vacancies start to form after the presence of jogs. As the jogs move, trails of vacancies are generated behind them. Similar results have been reported for MD simulations (Zhou et al., 1999).

In the above test, the dislocation density vectors of the intersecting parts are $\boldsymbol{\rho}^{(1)}=(0,\rho^{(1)},0)$ and $\boldsymbol{\rho}^{(2)}=(-\rho^{(2)},0,0)$ and the corresponding dislocation velocity vectors are $\mathbf{v}^{(1)}=(0,0,v)$ and $\mathbf{v}^{(2)}=(0,0,-v)$. The Burgers vectors are $\mathbf{b}^{(1)}=(0,b,0)$ and $\mathbf{b}^{(2)}=(b,0,0)$. According to Eq. (20), the jogs generated during a time increment $\Delta t$ are $\Delta\boldsymbol{\rho}^{(1)}_{\mathrm{jog}}=(-2vb\rho^{(1)}\rho^{(2)}\Delta t,0,0)$ and $\Delta\boldsymbol{\rho}^{(2)}_{\mathrm{jog}}=(0,-2vb\rho^{(1)}\rho^{(2)}\Delta t,0)$. Based on Eq. (24), both $\mathbf{b}^{(1)}\cdot(\boldsymbol{\rho}^{(1)}_{\mathrm{jog}}\times\mathbf{v}^{(1)})$ and $\mathbf{b}^{(2)}\cdot(\boldsymbol{\rho}^{(2)}_{\mathrm{jog}}\times\mathbf{v}^{(2)})$ have positive sign and so all jogs produce vacancies. A similar analysis shows that changing the direction of only one of the dislocation loops makes the jogs on both loops produce interstitials.

Figure 5 shows the evolution of the average dislocation density, jog density and vacancy concentration in the domain. Dislocation density increases initially due to the expansion of the loops. Then it decreases because the loops reach the periodic boundaries and self-annihilate with their images. The initial values of jog densities are zero, and they start to increase at about $t$ = 4 ns when the two dislocation loops first begin to intersect. The jog densities continue to increase until $t$ = 16 ns when the two dislocation loops completely pass each other. Then the jog densities remain nearly constant as the separated jogs only move in space. The small decrease in jog density is attributed to numerical diffusion. The evolution of the dislocation and jog densities is identical because of the identical initial conditions and speeds on both slip systems. The vacancy concentration increases after jogs are formed. When no more jogs are formed and the separated jogs move with a constant velocity, the vacancy concentration increases linearly over time.

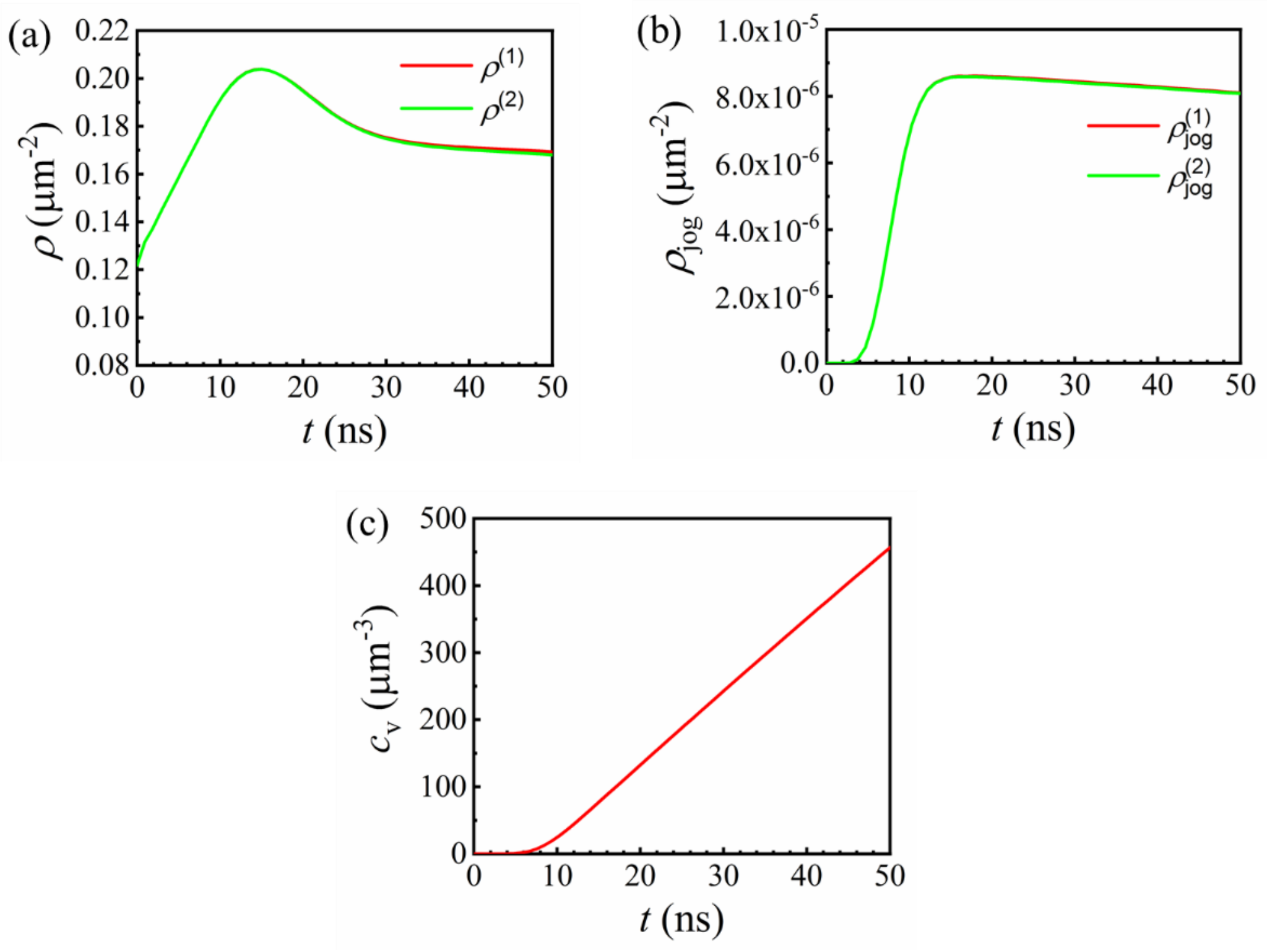

Figure 5. Average dislocation density, jog density and vacancy concentration as functions of time.

## 6.2 Evolution of expanding dislocation loop with jogs

A simple test problem is solved in this section to show the effect of the drag stress due to jogs on the evolution of a dislocation loop. This problem involves a dislocation loop with a radius of 1μm placed at the center of the simulation volume, as shown in Figure 6. The size of the simulation volume is 5μm×5μm×5.303μm, with edges along $[110]$, $[\bar{1}10]$ and $[001]$ crystallographic directions. The dislocation loop is on the $(111)$ slip plane with its Burgers vector along the $[\bar{1}10]$ direction. Initial jogs are assigned to the screw sides of the dislocation loop, and the jog density corresponds to cutting the dislocation loop by dislocation lines on slip plane $(\bar{1}11)$ with Burgers vector [110] (along *x*-axis), as shown in Figure 6(b). The jog density vector $\boldsymbol{\rho}_{\mathrm{jog}} = (\rho_{\mathrm{jog}}, 0, 0)$ has only one nonzero component, whos magnitude is proportional to the dislocation density described by Gaussian distribution around the screw parts of the dislocation loop with a scale factor of $5\times10^{-4}$. Material parameters are shown in Table 1 and Table 2. These parameters correspond to stainless steel (Ghoniem and Cho, 1979; Surh et al., 2004).

| | | | |
|---|---|---|---|
| Young's modulus (GPa) | 189 | Burgers vector *b* (nm) | 0.254 |
| Poisson's ratio | 0.26 | Drag coefficient (Pa.s) | $7.12\times10^{-6}$ |

Table 1. Material properties.

| | | | |
|---|---|---|---|
| Vacancy formation energy (eV) | 1.60 | Relaxation volume of a vacancy | -0.2Ω |
| Interstitial formation energy (eV) | 4.08 | Relaxation volume of an interstitial | 1.5Ω |
| Atom volume Ω ($m^3$) | $1.18\times10^{-29}$ | | |

Table 2. Parameters for point defect generation.

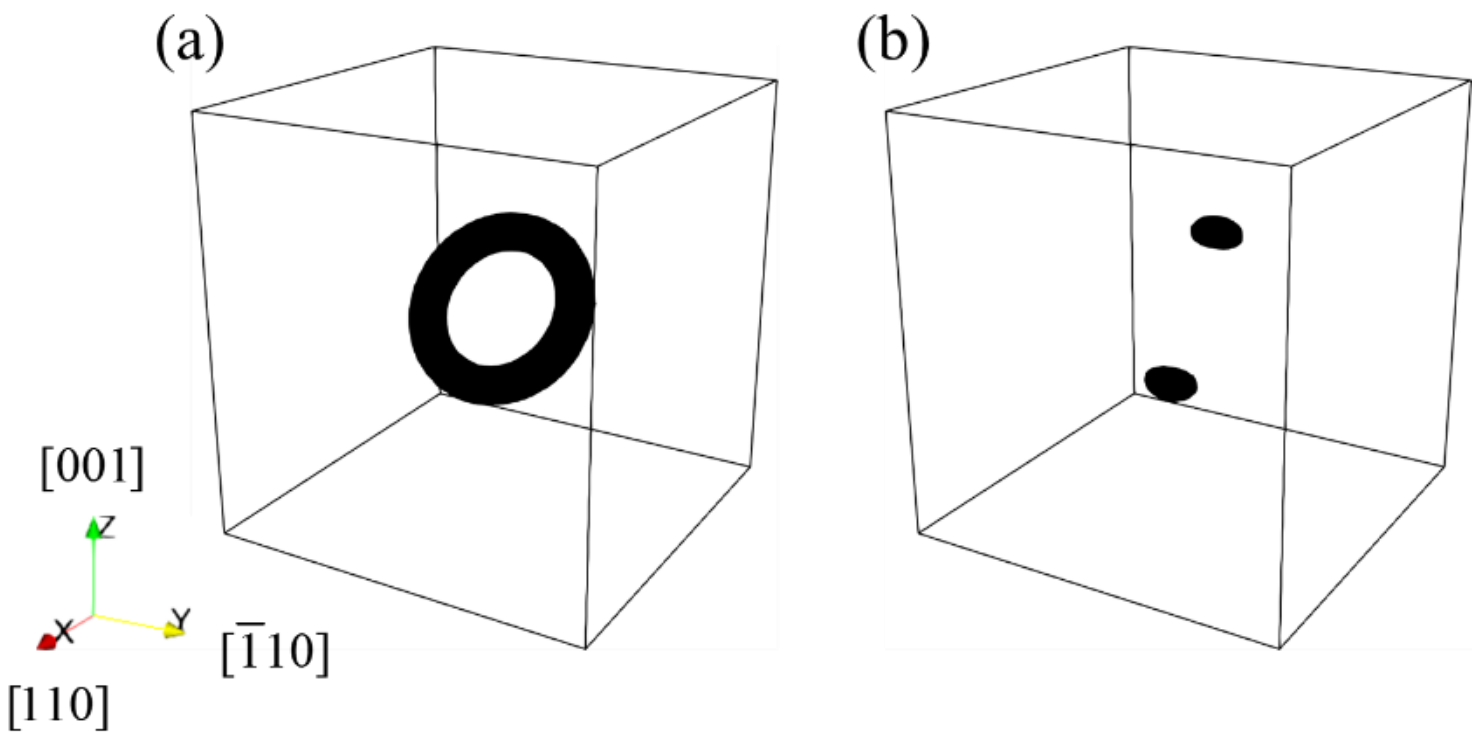

Figure 6. (a) A dislocation loop is placed on $(111)$ plane with Burgers vector along $[\bar{1}10]$ direction. (b) Jog density is assigned on the screw part of the dislocation loop.

An external load is applied to make the loop expand. The external load results in a resolved shear stress of 50 MPa on the dislocation loop. It is obvious that if there were no jogs, the dislocation loop would retain its initial circular shape during its expansion. However, by considering jog drag and point defect generation, the dislocation loop was found to evolve quite differently. This difference is shown below in Figure 7. As can be seen in the figure, the dislocation velocity is reduced where the jogs exist. The behavior is similar to that of a dislocation pinned by an obstacle; the dislocation line bows out near the parts where there are jogs. From the evolution of the jog density, it can be seen that the distance between upper and lower parts increases as the dislocation loop expands, since the jogs are tied to the dislocation line. Also, the region in which the jog density is non-trivial becomes thinner as the dislocation line continues to bow out.

The motion of the jog density is non-conservative and so, according to Eq. (24), point defects are produced in conjunction with the jog motion. Since the dislocation velocities are opposite at the upper and lower parts of the dislocation loop, the upper part generates interstitials and the lower part generates vacancies. As the region of jog density becomes thinner during loop evolution, the shapes of the regions where the interstitials and vacancies are generated change accordingly. On the other hand, since the formation energy of interstitials is larger than that of vacancies, vacancies are more easily generated and as such there will be more vacancies than interstitials at the end of the simulation. The result is clearly shown in Figure 8. In consequence, the jogs that generate interstitials are seen to impede the motion of the dislocation line more than the jogs that generate vacancies.

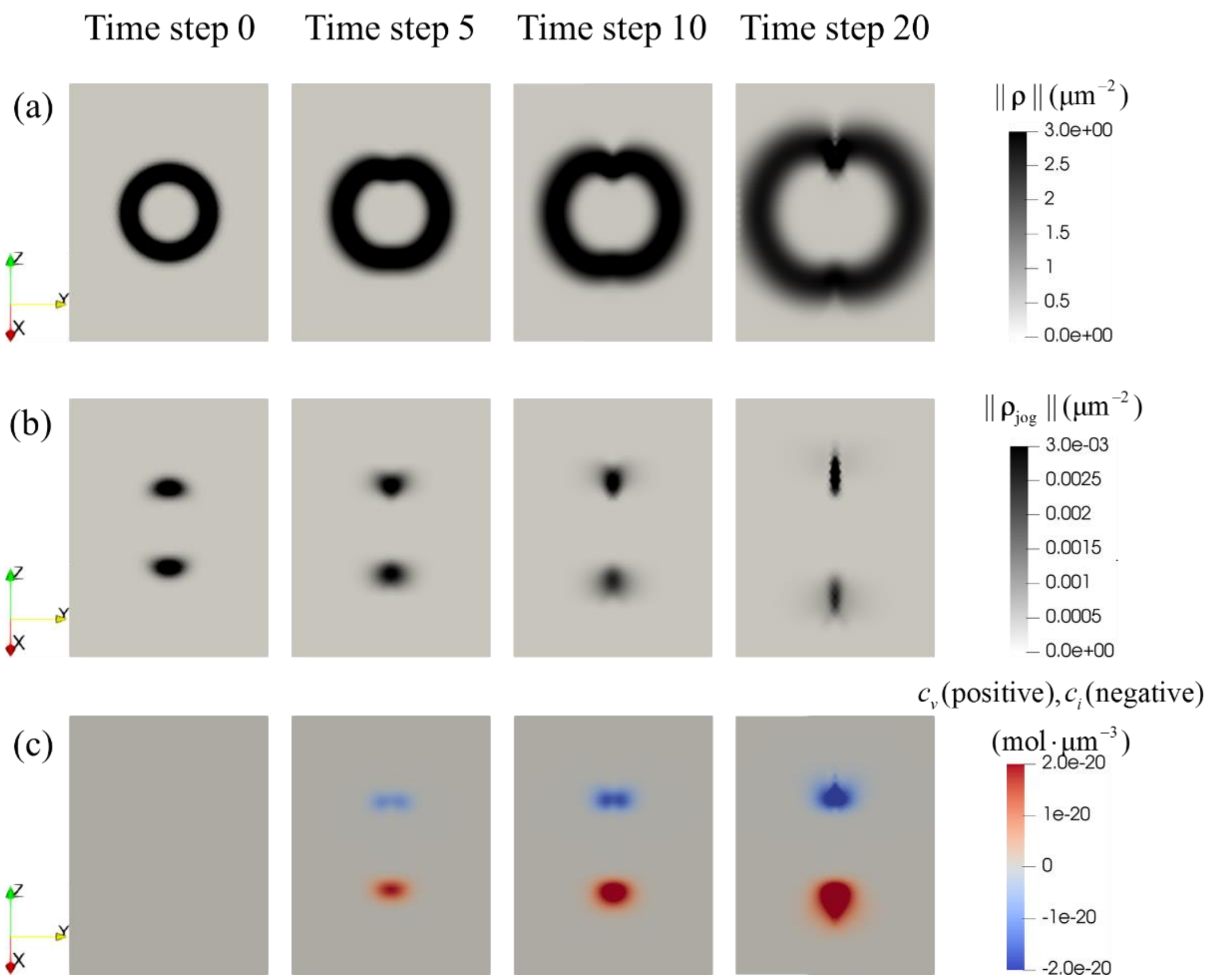

Figure 7. Dislocation loop expansion under the action of an external stress with jog drag on both the screw sides. Evolution of (a) dislocation density, (b) jog density, and (c) point defects (positive for vacancies and negative for interstitials).

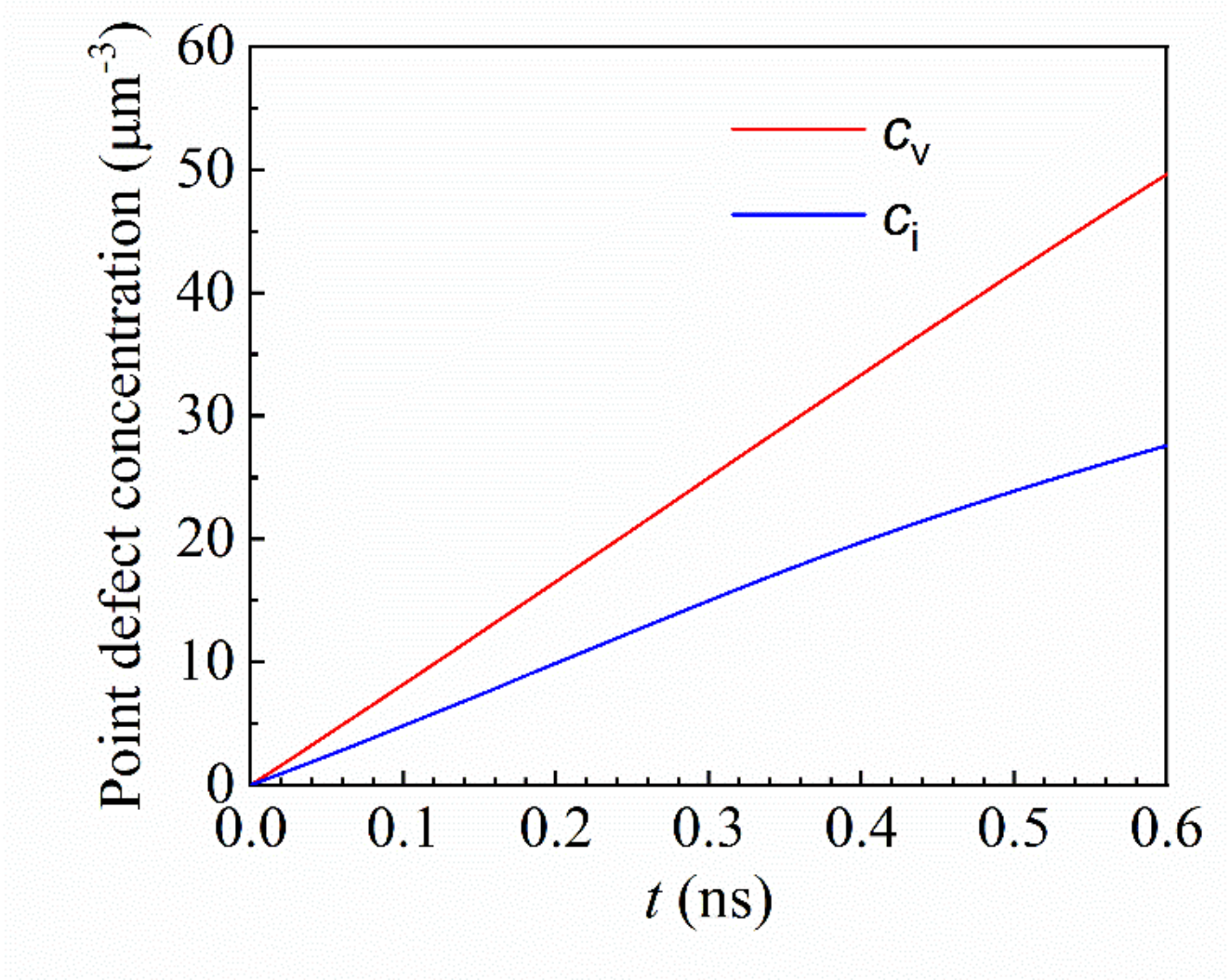

Figure 8. Evolution of the vacancy and interstitial concentration with time. More vacancies are generated than interstitials due to the lower formation energy of vacancies.

The jog drag stress and resolved shear stress corresponding to interstitial and vacancy generation are compared in Figure 9. It is obvious that the jog drag stress for generating interstitials is higher than the jog drag stress for generating vacancies. The resolved shear stress in Figure 9 includes the external stress (which is 50 MPa) and the stress due to eigen-strain induced by the dislocation loop under a periodic boundary condition. So one may as: under what circumstance will the drag stress have a prominent effect compared with the resolved shear stress? The drag stress expression, Eq. (43) can be rewritten as

$$\tau_{\mathrm{jd}}^{(k)} = \frac{E_{\mathrm{d}}}{\Omega} |\mathbf{s}^{(k)} \cdot (\boldsymbol{\xi}_{\mathrm{jog}}^{(k)} \times \boldsymbol{\eta}^{(k)})| \frac{\rho_{\mathrm{jog}}^{(k)}}{\rho^{(k)}} \tag{46}$$

where $\mathbf{s}^{(k)}$, $\boldsymbol{\xi}_{\mathrm{jog}}^{(k)}$ and $\boldsymbol{\eta}^{(k)}$ are unit vectors along the Burgers vector, jog direction, and dislocation velocity, respectively. $|\mathbf{s}^{(k)} \cdot (\boldsymbol{\xi}_{\mathrm{jog}}^{(k)} \times \boldsymbol{\eta}^{(k)})|$ is always less than or equal to unity. Regardless of these orientations, the maximum drag stress is $E_{\mathrm{d}}\rho_{\mathrm{jog}}^{(k)} / \Omega\rho^{(k)}$. So, for a given material, the ratio of jog density to dislocation density determines the magnitude of drag stress. For the material properties used in this simulation, taking the vacancy formation energy as an example, $E_{\mathrm{d}} / \Omega = 2.17 \times 10^4$ MPa. So, to generate 10 MPa drag stress, $\rho_{\mathrm{jog}}^{(k)} / \rho^{(k)}$ should be at least $4.61 \times 10^{-4}$.

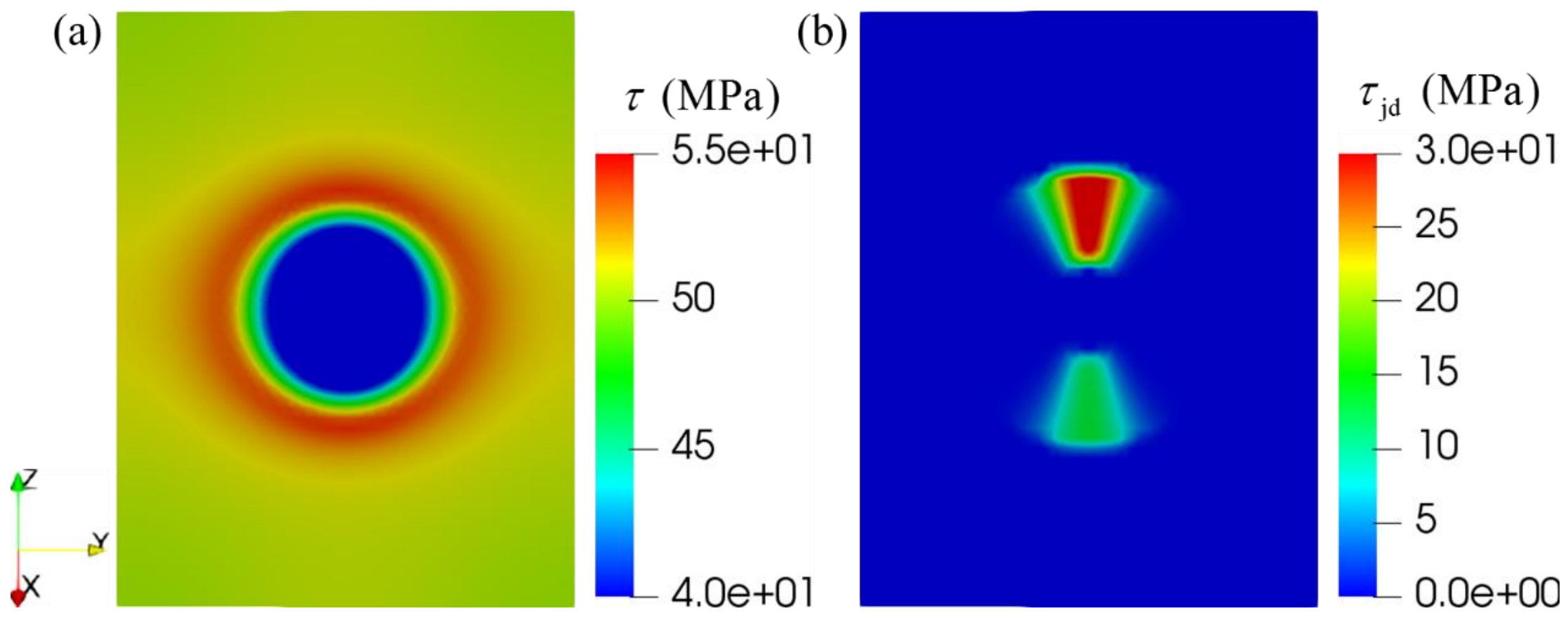

Figure 9. The (a) resolved shear stress and (b) jog drag stress at the first step of the simulation.

## 6.3 Effects of jog drag and defect generation on the mechanical response of single crystals

Two bulk crystal simulations were performed to study the mechanical response of a FCC crystal under uniaxial loading. In one simulation jogs and point defects are produced by the moving dislocations but the effects on dislocation motion of the eigen-strain of point defects and the drag stress of jogs were ignored. This model is referred to as the one-way coupling model. In the second simulation, the eigen-strain of point defects and drag stress of jogs were considered in dislocation dynamics, whereby point defects and dislocation jogs can also affect the evolution of dislocations. This model is referred to as the two-way coupling model.

The simulation domain is a $5\ \mu\text{m}\times 5\ \mu\text{m}\times 5.303\ \mu\text{m}$ box, discretized by a hybrid mesh of pyramid and tetrahedron elements. The mesh size is 62.5 nm. All 12 slip systems of FCC crystal are considered. The edges of the simulation domain are along $[110]$, $[\bar{1}10]$ and $[001]$ crystallographic directions. An initial dislocation density of $1.5\times 10^{12}\ \text{m}^{-2}$ is distributed across all 12 slip systems as loops. These loops have radii ranging from $2\ \mu\text{m}$ to $6\ \mu\text{m}$, and they are placed randomly in the domain with periodic boundary (the part of the loop exiting the domain will be re-entered from the opposite boundary). The initial dislocation configuration is shown in Figure 10. It is used in both the one-way and the two-way coupled simulations. The crystal is then loaded along the [001] direction with a strain rate of 20 $s^{-1}$. Periodic boundary conditions are applied on the six surfaces of the domain. Most material parameters are the same as in Table 1 and Table 2, except that a different drag coefficient was used, which is $B = 2.5\times 10^{-4}\ \text{Pa}\cdot\text{s}$. This value was obtained from recent molecular dynamic simulations at a temperature at $T$=573 K. Additional parameters regarding the diffusion equations of point defect are listed in Table 3.

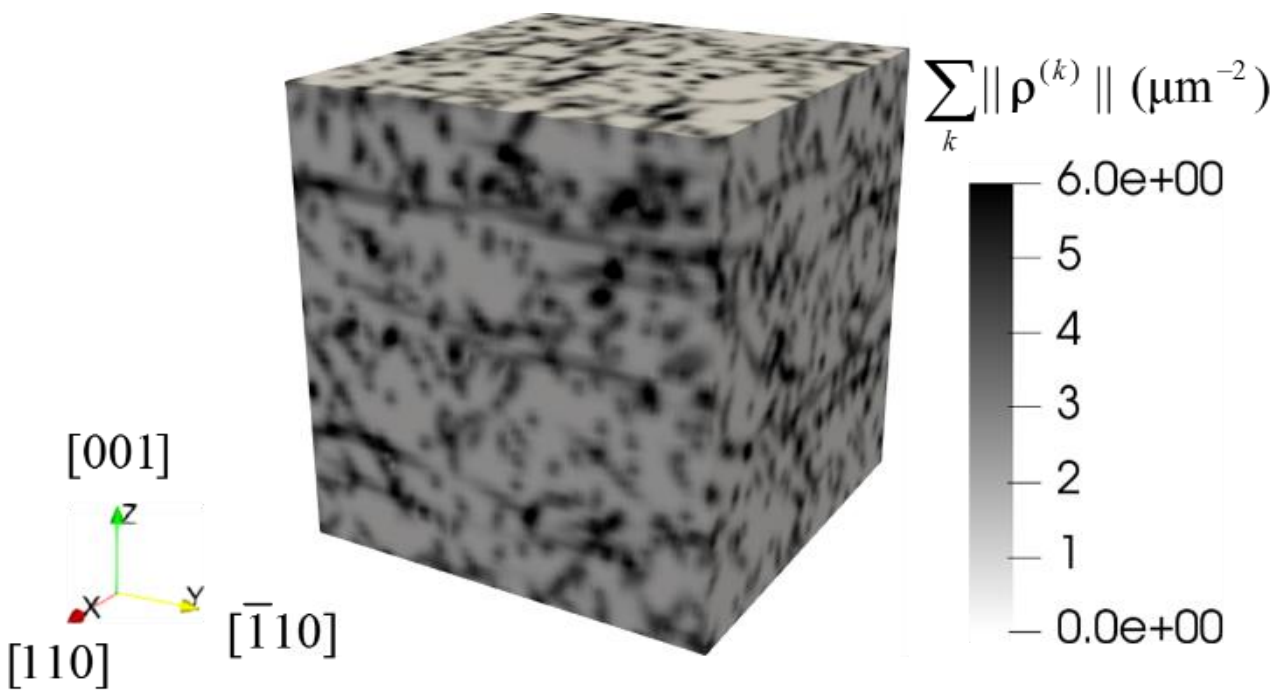

Figure 10. Initial dislocation configuration for the 3D bulk simulations.

| | | | |
|---|---|---|---|
| Gas constant (J/mol K) | 8.314 | Diffusivity of vacancy ($m^2/s$) | $3.28\times10^{-18}$ |
| Temperature (K) | 573 | Diffusivity of interstitial ($m^2/s$) | $5.76\times10^{-9}$ |
| Recombination coefficient ($m^3/s$) | $6.25\times10^{-17}$ | | |

Table 3. Parameters for point defect diffusion and recombination (Ghoniem and Cho, 1979; Surh et al., 2004).

The stress-strain curves and dislocation density evolutions of the one-way and two-way coupled simulations are shown in Figure 11. In part (b) of the figure, the scalar dislocation density, $\rho = \sum_k \| \boldsymbol{\rho}^{(k)} \|$, is displayed. It clearly shows that the two-way coupling between jogs, defects and dislocations results in a higher hardening rate than the one-way coupling (Figure 11(a)). There are two main reasons accounting for this difference. First, as the jog drag stress is considered in the two-way coupling, it is obvious that higher stress is required to move dislocations. Second, the dislocation density is also higher in the two-way coupling simulation (Figure 11(b)), which leads to stronger dislocation-dislocation interactions and a higher resistance to dislocation motion as per Taylor hardening Eq. (40). The higher dislocation density in the two-way coupled model can be explained by the result of the test simulation in section 6.2. The jog drag stress impedes the motion of part of the dislocation density thus forcing the dislocation lines to bow out and increase their length. This effect is similar to that of dislocations passing obstacles.

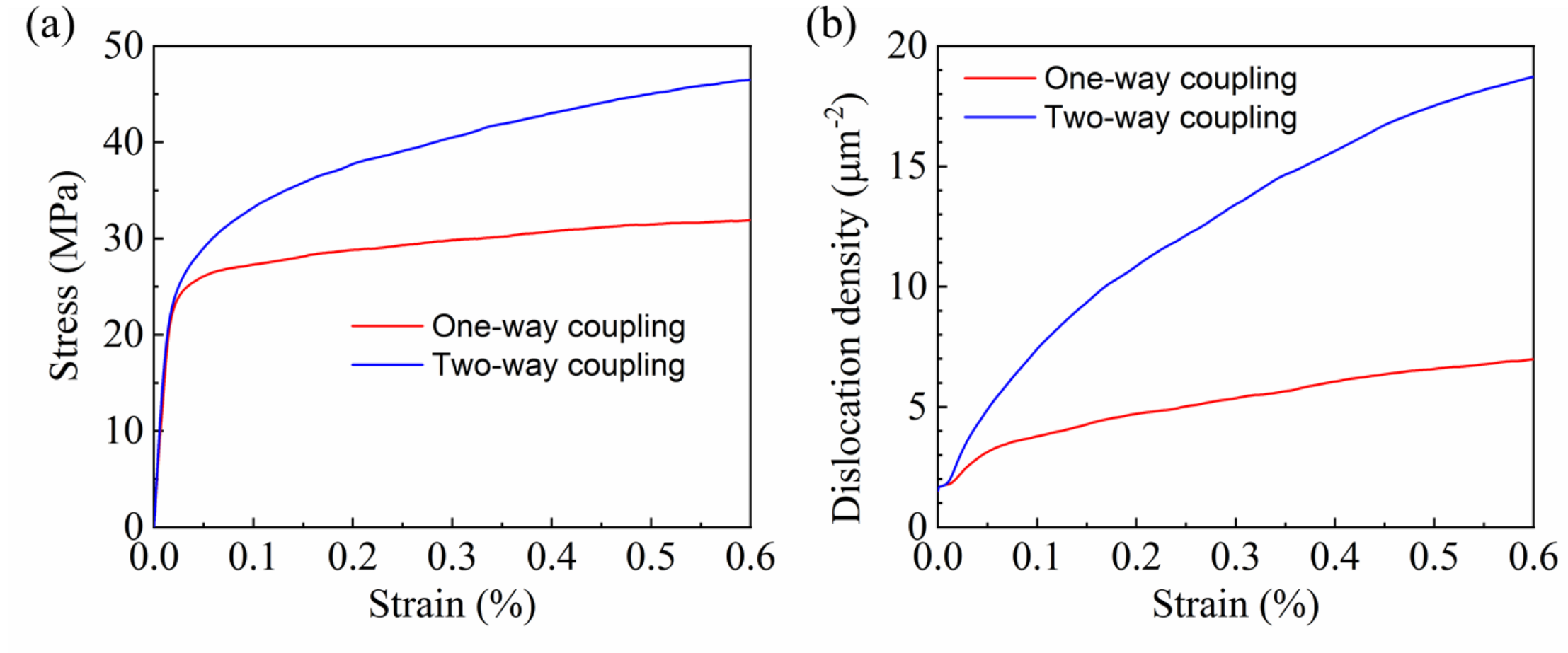

Figure 11. Stress-strain curves, (a), and the dislocation density versus strain, (b). Part (b) displays the scalar dislocation density, $\rho = \sum_k \| \boldsymbol{\rho}^{(k)} \|$.

The dislocation density evolutions on the individual slip systems for both the one-way and two-way coupled simulations are compared in Figure 12. The scalar density $\rho^{(k)} = \| \boldsymbol{\rho}^{(k)} \|$ is displayed in both parts of the figure. For the case of one-way coupling under [001] type loading, the density evolution on the eight active slip systems and the four inactive ones is clearly distinct. The dislocation densities on the active slip systems increase faster than their counterparts on the inactive ones. Also, within each group of slip systems, the densities are closer to one another due to the symmetry of the slip systems relative to the loading axis. In the case of two-way coupling, the dislocation densities on the inactive slip systems are relatively lower than the dislocation densities on the active slip systems as well. However, the density evolution shows significant differences among the slip systems, active or inactive. This difference is due to the random creation of jogs and the asymmetry of the jog drag stress.

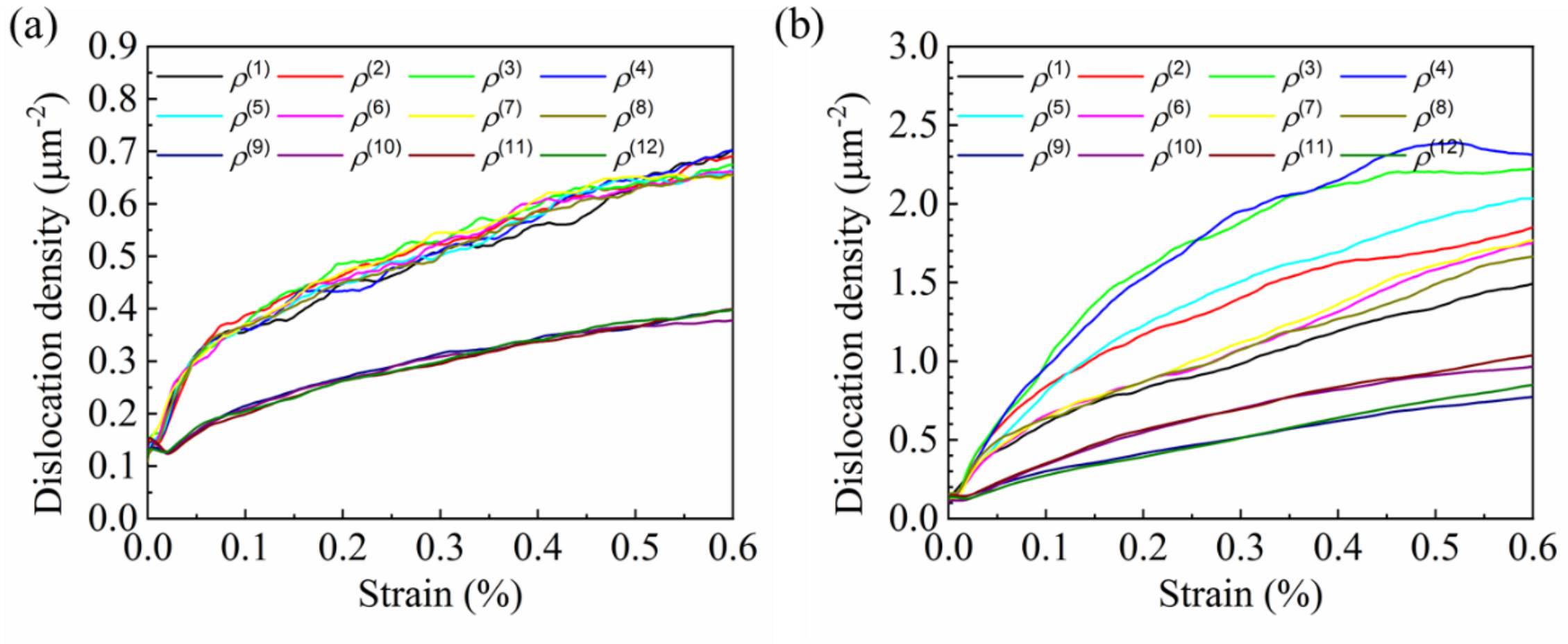

Figure 12. Dislocation density evolutions on individual slip systems. (a) One-way coupling case. (b) Two-way coupling case. The scalar dislocation density on each slip system, $\rho^{(k)} = \|\boldsymbol{\rho}^{(k)}\|$, is displayed.

Figure 13 shows the evolution of the jog density, $\rho_{\text{jog}} = \sum_k \|\boldsymbol{\rho}_{\text{jog}}^{(k)}\|$, in both the one-way and two-way coupling cases. Figure 13(a) shows that the jog density increases almost linearly with strain in both cases. However, the jog density in the two-way coupled simulation is much higher than its counterpart in the one-way coupled simulation. This difference is due to the feedback mechanisms between jog drag, the increase in the dislocation density, and the subsequent increase in the jog density due to more dislocation-dislocation cutting when the density is higher. The ratio of jog density to dislocation density is shown in Figure 13(b). This ratio also increases with strain and is higher in the case of two-way coupling than for the one-way coupling. It is important to mention here that the dislocation density is often proportional to the square root of strain or to the strain to some power less than 1, see Figure 11(b), but the jog density is proportional to strain itself as shown Figure 13(a). The latter dependence has to do with the fact that jogs are formed via the dislocation motion and intersection, i.e., with the accumulation of strain. From the strain dependence of both the dislocation density and the jog density, it can be concluded that the jog density is proportional to the square of dislocation density or to the density to some power higher than unity. This behavior is consistent with the jog density generation equation Eq. (20), in which the jog generation rate involves a density product. As discussed in section 6.2, the ratio of jog density to dislocation density determines the drag stress of jogs. According to the analysis there, a value of $\rho_{\text{jog}} / \rho = 2\times10^{-4}$ at 0.6% strain will result in an average jog drag stress of 4.34 MPa.

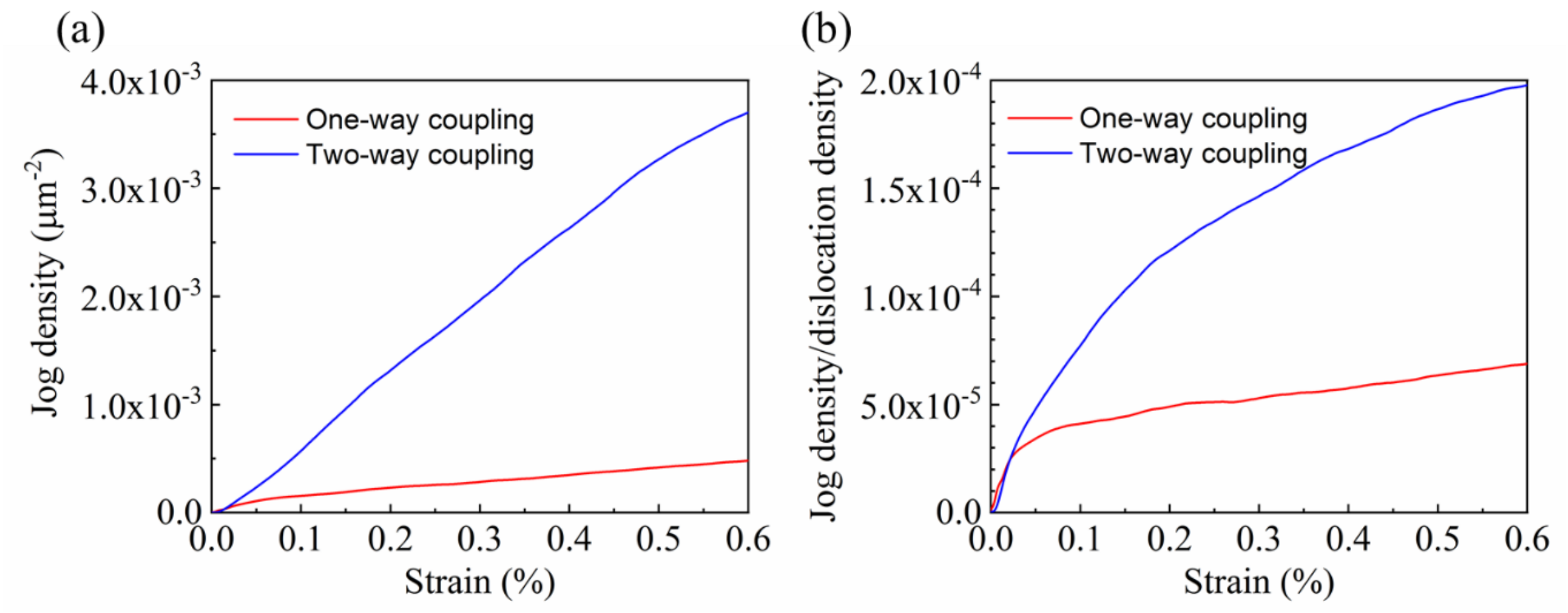

Figure 13. Jog density evolution. (a) Jog density versus strain, (b) The ratio of jog density to dislocation density versus strain.

Figure 14 shows the evolution of point defects. For the one-way coupled simulation, since point defects and jogs have no effect on dislocations, the increase of point defect density follows a constant trend – nearly following the jog density. For the two-way coupled simulation, the evolution of point defect densities exhibits two regimes. Initially, point defect density increases with the same rate as in the one-way coupled simulation, since the jog density is low and the jog drag stress is not sufficiently high to affect the dislocation velocity. Beyond 0.1% strain, as the ratio of jog density to dislocation density reaches about $8\times10^{-5}$, the generation rate of point defect decreases significantly. Although the jog density in the two-way coupled simulation is much higher than the jog density in the one-way coupled simulation, there are far fewer point defects generated in the two-way coupled simulation. This means that jogs have a much smaller average velocity in the two-way coupled simulation even though their density is higher. On the other hand, from theory, it is expected that the difference in defect formation energy will create an asymmetry of drag on dislocations resulting in the production of more vacancies than interstitials. In the one-way coupling simulation, the effect of jog drag stress on dislocations is not accounted for, and hence the difference in formation energies of different point defects does not affect their production, which in this case only depends on the densities of jogs producing the two types of defects. The bias due to initial configuration of dislocations favors the generation of interstitials. In the two-way coupling simulation, although there are more interstitials in the beginning, vacancies are generated faster than interstitials beyond about 0.35% strain. This would have important implications as far as void nucleation is concerned and its role in ductile fracture of metals.

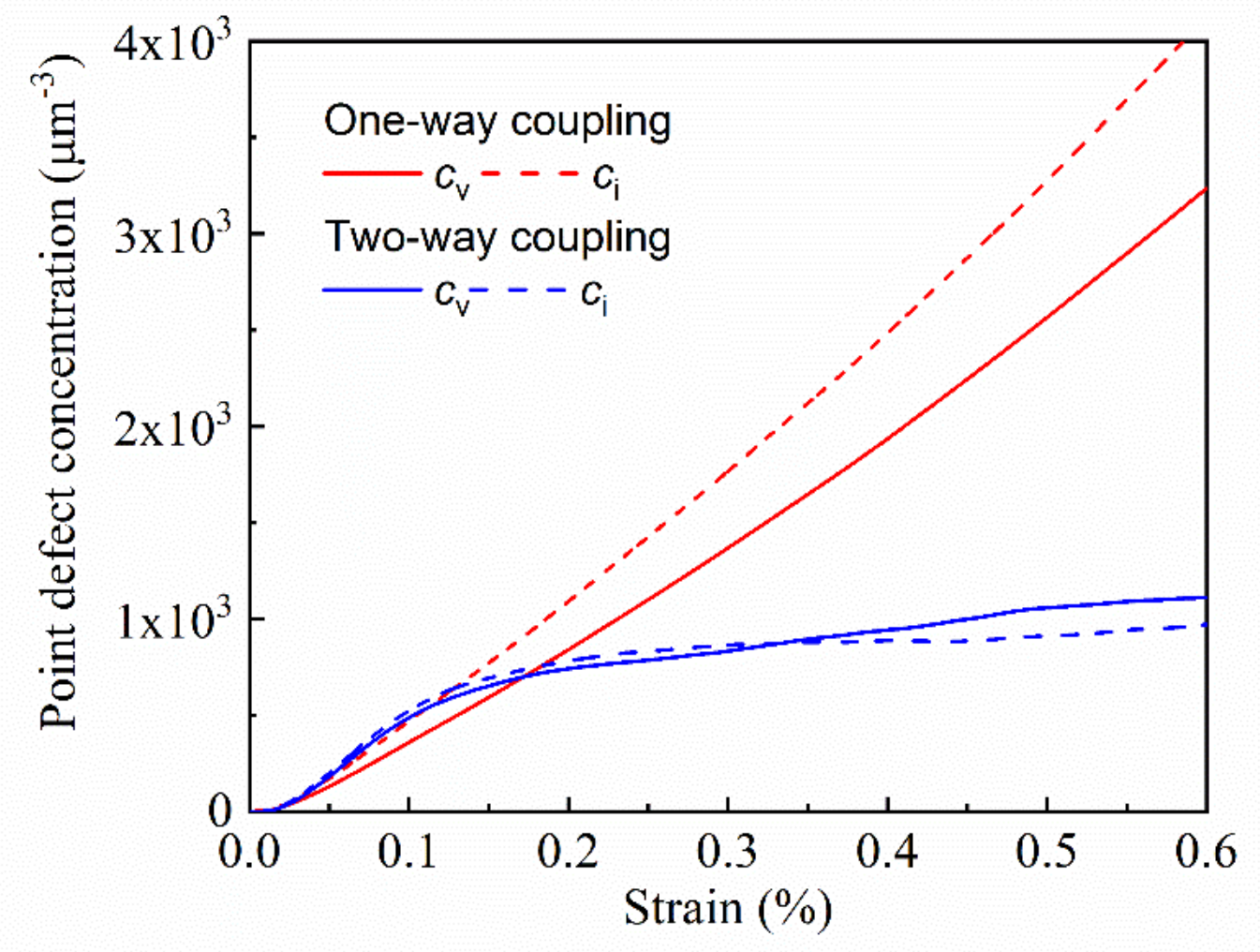

Figure 14. Evolution of the vacancy and interstitial concentration in the one-way and two-way coupled simulation of dislocations, jogs and point defects.

Figure 15 compares the dislocation microstructure for the two simulations. The dislocation density is shown on a (111) slip plane at 0.6% strain for the one-way coupling in Figure 15(a) and the two-way coupling in Figure 15(b). In both cases, the heterogeneity of the dislocation density is anticipated. Dislocations are likely to accumulate in a pattern along three specific directions. One preferred direction is horizontal and the other two are oriented $\pm\pi/3$ from the horizontal line. These directions are the intersections of the slip plane with the other three slip planes of the FCC crystal. The dislocation density pattern is more obvious in the two-way coupled case. As shown in Section 6.2, the drag stress of jogs tends to retard dislocation motions, and the dislocation density will increase as dislocations bow out more. The intersections of the slip plane are the places where jogs are more likely to be formed. For this reason, in the two-way coupled case, localized dislocation densities of a larger magnitude are more likely.

The vacancy concentration on the same slip plane are shown in Figure 16 for the one-way and two-way coupled cases. The vacancy pattern appears to have a finer structure in the case of one-way coupled simulation than in the two-way coupled case. In the one-way coupling case the drag stress of dislocation jogs is not considered. Once the jogs are formed by dislocation intersections they move together with dislocations. As dislocation intersections happen everywhere in the domain, point defects can also be generated at the places where jogs have moved. For the two-way coupling case, to generate point defects, the local resolved shear stress must be large enough to overcome the drag stress caused by the jogs. For the locations where the stress does not satisfy this criterion, the jogs are not able to move. Consequently, fewer point defects are generated and the resulting deformation patterns exhibit larger wavelengths on average.

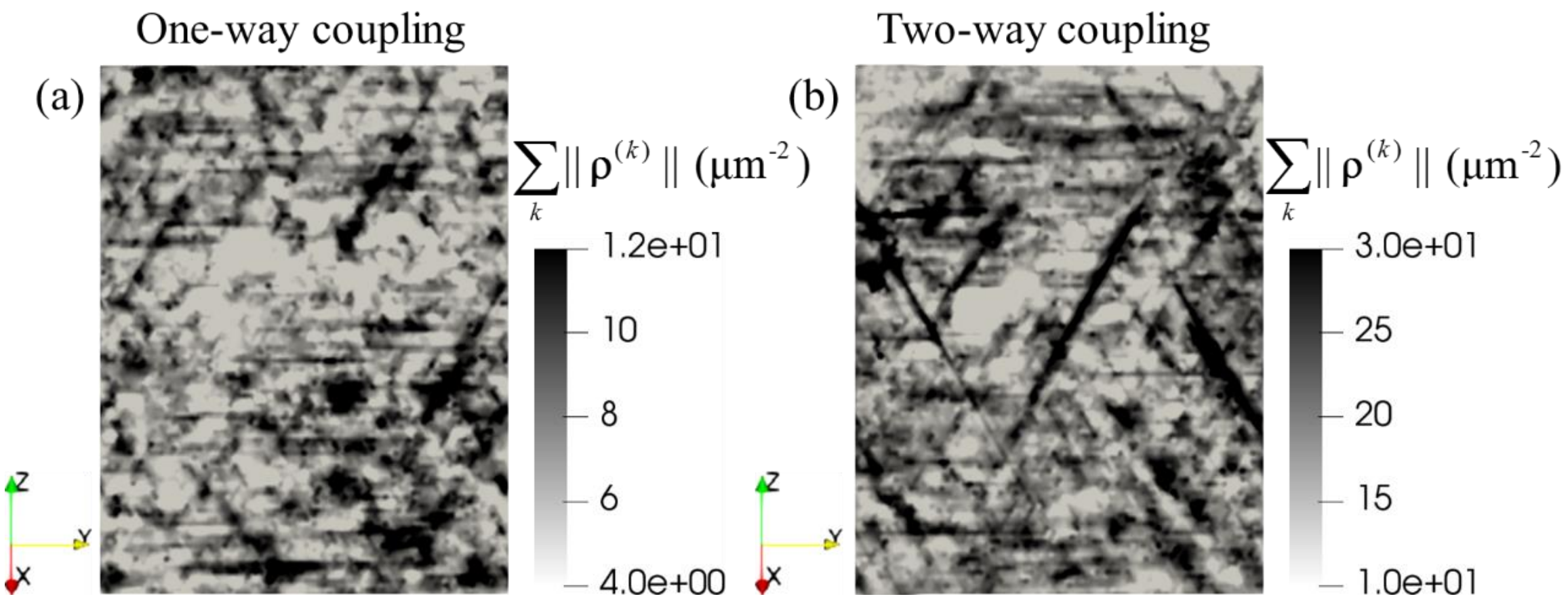

Figure 15. Dislocation density on (111) slip plane at 0.6% strain. (a) One-way coupled simulation, and (b) two-way coupled simulation.

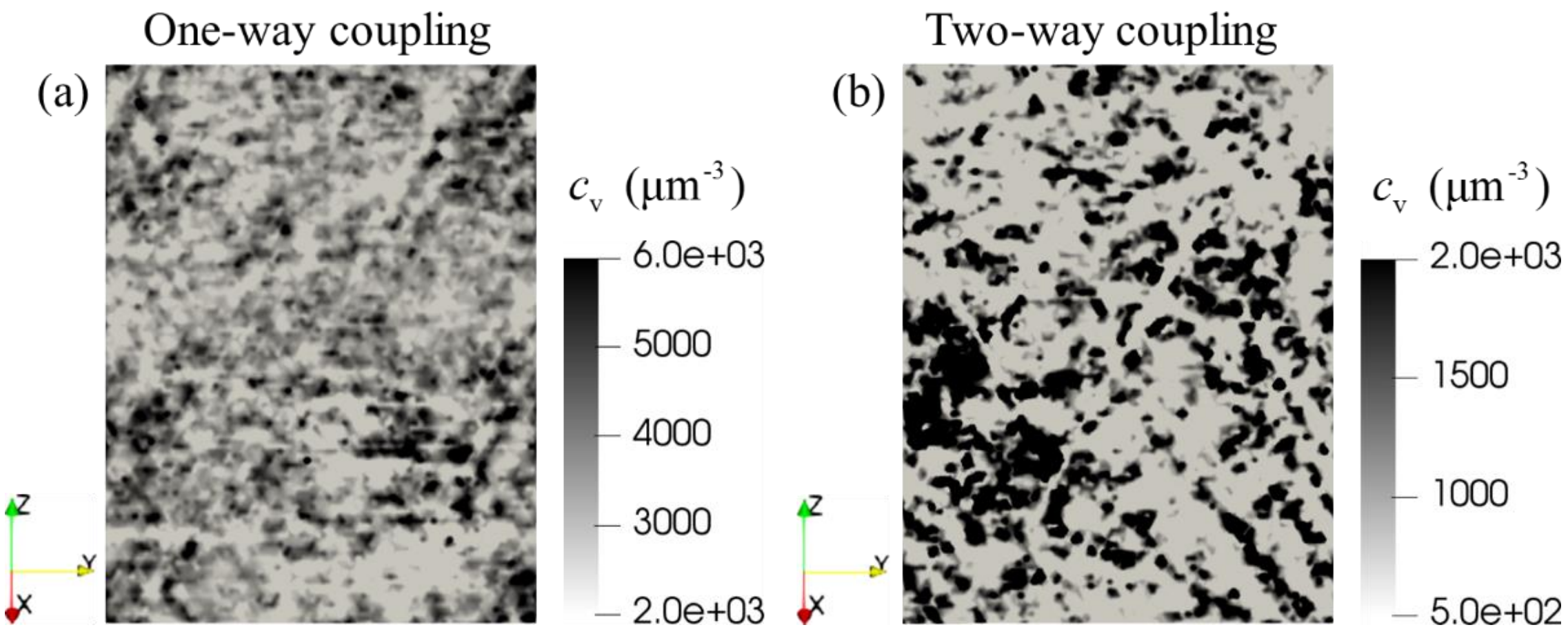

Figure 16. Vacancy concentration on (111) slip plane at 0.6% strain. (a) One-way coupled simulation, and (b) two-way coupled simulation.

In in the coupled continuum dislocation dynamics/point defect generation model developed and solved in this work, jog formation and transport serve as the mechanism of point defect generation and drag on dislocations. In our formulation, the line direction of dislocations and jogs are considered as well as the Burgers vectors, as in Eqs. (20) and (24). It is usually not possible to do this in crystal plasticity at the mesoscale, since dislocations are represented by scalar dislocation densities. Even in discrete dislocation dynamics, modelling of the jog formation and evolution is a challenging and cumbersome exercise since it requires the tracking of the individual jogs and perhaps also the individual point defects generated. The test problem presented in Section 6.1 clearly shows how effective continuum modeling of dislocations is in describing the process of point defect generation by moving dislocations, a new advantage offered by the current development. Dislocation jogs are formed first by intersecting dislocation lines. As dislocation jogs move, trailing point defects are generated along their path. This is exactly what we expect from the proposed model. On the other hand, the existence of jogs can change the evolution of dislocations due to drag stress. The drag stress is carefully formulated by considering the energy consumption for the type of point defect generated. As shown in the example in Section 6.2, jogs are similar to obstacles that impede the motion of dislocations, leading to the bow out of dislocations. This fact implies that, by considering the effect of jogs, the length of the dislocation line will increase, resulting in a higher dislocation density. Furthermore, a higher dislocation density will contribute more to strain hardening. This conclusion is confirmed by the bulk simulation in Section 6.3, as shown in Figure 11. Another important observation is that the asymmetry in the jog drag stress for jogs oriented for vacancy versus interstitial generation leads to an asymmetry of the rate of the point defects generated, essentially leading to a higher vacancy generation rate than for interstitials, which shows in the cumulative average defect concentrations in the deformed crystal. This asymmetry explains why there are often more vacancies than interstitials in plastically deformed metals. The current results demonstrate that the model presented here may help to understand void nucleation during plastic deformation.

## 7 Concluding remarks

To summarize, a model for point defect generation by the jog formation and transport mechanism during plasticity is proposed within the framework of vector density-based continuum dislocation dynamics. As a part of this model, detailed equations for jog formation and transport were developed as a part of continuum dislocation dynamics. Jogs were assumed to form as a result of intersection of dislocations on various slip systems. Jogs were also assumed to be quasi-particles attached to dislocations that move together with the dislocations with the same speed and direction. The rate of point defect generation associated with jog transport was formulated in terms of the volume non-conservation associated with jog motion, i.e., with the non-glide part of the motion of the dislocations in the crystal. Balance equations for the vacancies and interstitials including their rate of generation due to jog transport, recombination and diffusion were also established. The effect of point defects on dislocations was further included via the stress induced by their eigen-distortion. Finally, a jog drag stress was introduced into the mobility law of dislocations thus accounting for the energy expended in producing the point defects. All model equations were coded in conjunction with continuum dislocation dynamics using the finite element method and implicit time integration. Test problems were presented, including jog formation and transport and the associated vacancy and interstitial generation, and the effect of jog drag stress on the dislocation evolution. Coupled solutions of the plasticity and point defect generation problem under uniaxial load were presented.

The results show that fully coupled dislocation and point defect dynamics via jog drag results in a higher dislocation density and a higher hardening rate. The results also show that the asymmetry of jog drag stress between jogs oriented for vacancy versus interstitial generation leads to higher vacancy generation and accumulation past the initial straining of the crystal. The dislocation and point defect patterns are also found to exhibit longer wavelengths in the case of fully coupled dislocation and point defect dynamics. The model as it currently stands is considered a first step toward generalized defect dynamics modeling of plasticity of metals in which point defect generation plays an important role, e.g., in situations involving hydrogen effects, and in cases where ductile fracture via void formation is important.

## Acknowledgements

The authors are grateful for the support from the Naval Nuclear Laboratory, operated by Fluor Marine Propulsion, LLC for the US Naval Reactors Program. A. El-Azab assisted with the theoretical formulation of the problem with support from the US Department of Energy, Office of Science, Division of Materials Sciences and Engineering, through award number DE-SC0017718 at Purdue University.